\documentclass[12pt,english,floatfix,superscriptaddress,aps,prd,preprint,showkeys,nofootinbib]{revtex4}
\usepackage{subcaption}
\usepackage{tikz}
\usepackage{multirow}
\usetikzlibrary{decorations.pathmorphing}
\usetikzlibrary{quotes,angles}
\usepackage{amsmath}
\usepackage{amssymb}
\usepackage{amsbsy}
\usepackage{amsfonts}
\usepackage{amsopn}
\usepackage{amstext}
\usepackage{graphicx}
\usepackage[english]{babel}
\usepackage{color}
\usepackage{slashed}
\usepackage{esint}
\usepackage[dvips]{epsfig}
\usepackage[dvips]{graphicx}
\usepackage{float}
\usepackage{units}
\usepackage{textcomp}
\usepackage{wasysym}
\usepackage{hyperref}
\usepackage{slashed}
\usetikzlibrary{arrows} 
\usetikzlibrary{decorations.markings}

\begin{document}

\title{Fermionic greybody factors and strong gravitational lensing by Lorentz-violating global monopole}
\author{Fernando M. Belchior}
\email{belchior@fisica.ufc.br}
\affiliation{Departamento de F\'isica, Universidade Federal do Cear\'a,\\ Campus do Pici, 60455-760, Fortaleza, Cear\'a, Brazil.}
\author{Roberto V. Maluf}
\email{r.v.maluf@fisica.ufc.br}
\affiliation{Departamento de F\'isica, Universidade Federal do Cear\'a,\\ Campus do Pici, 60455-760, Fortaleza, Cear\'a, Brazil.}

\author{Ana R. M. Oliveira}
\email{ana.rafaely@academico.ufpb.br}
\affiliation{Departamento de F\'isica, Universidade Federal da Para\'iba,
 58051-970, João Pessoa, Para\'iba, Brazil}

\author{Albert Yu. Petrov}
\email{petrov@fisica.ufpb.br}
\affiliation{Departamento de F\'isica, Universidade Federal da Para\'iba,
 58051-970, João Pessoa, Para\'iba, Brazil}
\author{Paulo J. Porf\'irio}
\email{pporfirio@fisica.ufpb.br}
\affiliation{Departamento de F\'isica, Universidade Federal da Para\'iba,
 58051-970, João Pessoa, Para\'iba, Brazil}

\begin{abstract}
In this work, we study the greybody factors (GFs) of spin 1/2 and spin 3/2 fermions for a black hole with global monopole in self-interacting Kalb-Ramond gravity with Lorentz symmetry violation. For our purpose, we consider the Dirac and Rarita-Schwinger equations in curved spacetime by proceeding with separating these equations into sets of radial and angular equations. Using the analytical solution of the angular equation, the Schr\"{o}dinger-like wave equations with potentials are derived by decoupling the radial wave equations using the tortoise coordinate. Moreover, we calculate the angular deflection of light in the strong field limit. With the expression for angular deflection in the strong field limit, we compute the positions as well as magnification of the respective relativistic images. We compute the shadows cast by the Lorentz-violating (LV) black hole with a global monopole and analyze how the LV parameter and the monopole charge affect the shadows.

\end{abstract}

\maketitle

\section{Introduction}

Understanding the quantum nature of black holes requires the General Relativity (GR) to be corrected. In this context, modification of GR as $f(R)$ and $f(R, T)$ had become an attractive research line \cite{Sotiriou:2008rp, Clifton:2011jh, Akbar:2006mq, delaCruz-Dombriz:2009pzc, Harko:2011kv}. Such theories are constructed in the spirit of Riemannian geometry, with the metric as a dynamical variable and vanishing torsion. Starting from the Einstein-Hilbert action, it is possible to build these extensions by writing a generalized action as a function of the curvature scalar $R$ and others invariants such as momentum-energy tensor $T$ or Ricci contraction $R_{\mu\nu}R^{\mu\nu}$. Researchers have investigated such theories both within the metric formalism and the metric-affine one (the Palatini formalism) \cite{Amarzguioui:2005zq, Koivisto:2007sq, Olmo:2011uz}.

The topic of Lorentz symmetry breaking has become a significant key area of research, covering various theoretical frameworks \cite{Kostelecky:1988zi, Colladay:1996iz, Colladay:1998fq, Carroll:1989vb}. One prominent approach to study this subject involves the use of vector or tensor fields with a non-trivial potential so that these fields can assume a non-zero vacuum expectation value, leading to spontaneous Lorentz symmetry breaking (LSB). Recently, researchers have focused on an antisymmetric tensor field known as Kalb–Ramond (KR) field \cite{Altschul:2009ae, Maluf:2018jwc, Assuncao:2019azw, Aashish:2019ykb}. This field, known to emerge in the context of string theory, provides a significant tool to scrutinize the features of LSB. A next step in studying the impacts of the KR field consists of considering the corresponding Lorentz-violating (LV) effects within the gravitational scenario.

The study of black holes (BH) and wormholes in the context of Kalb–Ramond (KR) gravity has gained considerable prominence in recent works \cite{Maluf:2021ywn, Lessa:2019bgi, Yang:2023wtu, Ovgun:2025ctx}, where the KR field can be non-minimally coupled to gravity along with a smooth self-interaction potential, leading to spontaneous LSB. As a consequence, an exact solution describing a static, spherically symmetric black hole has been found \cite{Lessa:2019bgi, Yang:2023wtu}. Following these seminal works, researchers have examined some physical properties, such as the dynamics both of massive and massless particles, gravitational deflection of light, and the shadows cast by Schwarzschild-like black holes \cite{Filho:2023ycx, Junior:2024vdk, Junior:2024ety, Jumaniyozov:2024eah}. Furthermore, there have been papers focused on charged black holes \cite{Duan:2023gng, Jumaniyozov:2025wcs, al-Badawi:2024pdx}, neutral \cite{Liu:2024oas} and rotating black holes \cite{Kumar:2020hgm, Zubair:2023cor,Kuang:2022xjp}. More recently, a study on black hole with global monopole was developed \cite{Belchior:2025xam}, where classical tests in the solar system and scattering, absorption and greybody factors of spin 0 particles were treated. 

A notable topic in the context of black holes is the greybody effect, which
refers to the radiation emitted as a result of the Hawking temperature. The greybody bound signifies the maximum deviation of this radiation from the ideal blackbody spectrum. In particular, one can investigate greybody effect for both bosonic and fermionic fields \cite{Maldacena:1996ix, Klebanov:1997cx, Boonserm:2023oyt, Boonserm:2021owk, Al-Badawi:2023emj}. Some papers have addressed this topic in the context of LV theories \cite{Guo:2023nkd, Al-Badawi:2023xig, Singh:2024nvx, Belchior:2025xam}.

Recently, the study of gravitational lenses has become an important field of research in the cosmological context \cite{Bozza:2002zj, Tsukamoto:2016jzh, Cunha:2018acu, Gyulchev:2008ff}. In this conjecture, gravitational lenses enable researchers to investigate the distribution of structures, dark matter, and some other topics \cite{James:2015yla, Furtado:2020puz, Nascimento:2020ime, Tsukamoto:2020bjm}. There are papers addressing gravitational lenses in both weak and strong field  regimes. This topic has been investigated in the context of LV theories in recent works \cite{Junior:2024vdk, Gao:2024ejs, Kuang:2022xjp, Ovgun:2018ran, Filho:2024isd}.

In this paper, we pursue two aims. The first one is to obtain the effective potential and compute the greybody factor for spin 1/2 and spin 3/2 fermions. The second one is to calculate the angular deflection of light by considering the strong field limit. From the angular deflection of light, one investigates the observables related to the gravitational lensing and also the shadows cast by the black hole solution within the KB theory using optically thin ring/shell model.

The paper is outlined as follows: in Section \ref{s2}, we present the metric that describes the black hole with a global monopole in KR gravity. In Section \ref{s3}, one obtains the Schr\"{o}dinger-like wave equations and the effective potentials for spin 1/2 and spin 3/2 fermions and computes the greybody factors. In Section \ref{s4}, we utilize the methodology developed by Bozza and Tsukamoto to obtain the expression for the angular deflection of light in the strong field limit. In Section \ref{s5}, we introduce an expression for a deflection of light in the strong field limit in the lens equation. In Section \ref{shadows}, we construct the shadow images of the LV black hole in the presence of a global monopole, and we study the departures from the standard (Schwarzschild) case. Finally, in Section \ref{s6}, we discuss the main results and present the conclusions of our research.

\section{Lorentz-violating global monopole}\label{s2}

We start our study introducing the LV black hole solution obtained in the presence of a global monopole for the purpose of determining the greybody factors of spin 1/2 and spin 3/2 fermions in the next sections and posterior study of gravitational lensing in the strong field regime. Therefore, the static and spherically symmetric black hole is described by the following line element \cite{Belchior:2025xam}
\begin{align}\label{metric}
    ds^2=-F(r)dt^2+\frac{dr^2}{F(r)}+r^2(d\theta^2+\sin^2{\theta}d\phi^2),
\end{align}
where the function $F(r)$ is given by
\begin{align}\label{sol1}
    F(r)=\frac{1-\kappa\eta^2}{1-\gamma}-\frac{2M}{r},
\end{align}
where we have defined the LV parameter $\gamma=\varepsilon\,\vert b\vert^2 /2$. From (\ref{sol1}), we obtain the radius of the event horizon given by
\begin{align}\label{cr}
  r_h=\frac{2M(1-\gamma)}{1-\kappa\eta^2},  
\end{align}
while the Hawking temperature reads
\begin{align}
 T_H=\frac{1}{8\pi} \frac{(1-\kappa\eta^2)^2}{(1-\gamma^2)^2}.  
\end{align}
The metric (\ref{metric}) will be used throughout this paper.

\section{Fermionic greybody factors}\label{s3}
Now, we will consider Dirac and Rarita–Schwinger equations in the LV global monopole spacetime.

\subsection{Spin 1/2 fermions}
Let us assume the action on curved space background for massive Dirac fermions as follows \cite{Boonserm:2021owk, Al-Badawi:2023emj, Al-Badawi:2023xig}
\begin{align}\label{action}
    S_{1/2}=\int d^4x\sqrt{-g}\,\overline{\psi}(i\Gamma^\mu D_\mu-m)\psi,
\end{align}
where $\Gamma^\mu=h_a\ ^\mu\gamma^a$ are the curved Dirac matrices related with the flat Dirac matrices $\gamma^a$.  Here, $D_\mu=\partial_\mu+\Omega_\mu$ denotes the covariant derivative, with $\Omega_\mu$ being the spin connection, namely
\begin{align}
\Omega_\mu=\frac{1}{4}\eta^{ab}h^c\ _\nu(\partial_\mu h_a\ ^\nu+\Gamma^\nu_{\mu\lambda}h_a\ ^\lambda)\gamma_c \gamma_b.    
\end{align}
The variation of the action (\ref{metric}) with respect to $\overline{\psi}$ gives the Dirac equation
\begin{align}
(i\Gamma^\mu D_\mu-m)\psi=0.    
\end{align}
For the metric (\ref{metric}), the vielbein can be defined as follows:
\begin{align}
 h^a\ _\mu=diag\bigg(\sqrt{F},\frac{1}{\sqrt{F}},r,r\sin\theta\bigg).   
\end{align}

After some lengthy manipulation (see Ref. \cite{Boonserm:2021owk} for more details), one obtains a one-dimensional
Schr\"{o}dinger-like wave equation with a Dirac field effective potential, namely,
\begin{align}
\frac{d^2 F_{\pm}}{d r^2_{\ast}} +(\omega^2-V_{\pm})F_{\pm}=0,   
\end{align}
where we have defined tortoise coordinates $r_{\ast}$ through the relation
\begin{align}
    \frac{dr_{\ast}}{dr}=\frac{2\omega (\lambda^2+m^2r^2)+\lambda m F  }{2\omega F(\lambda^2+m^2r^2)}.
\end{align}

Besides, we write the effective potentials for spin 1/2 fermions as follows:
\begin{align}
V_{\pm}=\pm\frac{d W}{d r_{\ast}}+W^2,    
\end{align}
where
\begin{align}
 W=\frac{\sqrt{F}}{r}\frac{2\omega (\lambda^2+m^2r^2)^{3/2}}{2\omega (\lambda^2+m^2r^2)+\lambda m F}.   
\end{align}
As we can see, the above potential is essentially dependent on the metric function and on the mass and the energy of the fermionic field. On the other hand, when setting $m=0$ in we obtain the following effective potential for massless Dirac fermions
\begin{align}\label{mlp}
  V_{\pm}=\frac{\lambda}{r^2}\bigg(\lambda F\pm\frac{rF^{\prime}}{2\sqrt{F}}\mp F^{3/2}\bigg),
\end{align}
where the prime indicates a derivative with respect to the usual radial coordinate $r$. Once we have obtained the effective potential, we can use it to compute the GF of Dirac fermions. To achieve such a goal, we will employ the general semi-analytic bounds, which allow us to perform a qualitative analysis of the results. As a starting point, we define the transmission probability 
\begin{align}
 \sigma(\omega)=\mathrm{sech}^2\bigg(\int_{-\infty}^{\infty}\rho(r_{\ast}) dr_{\ast}\bigg),   
\end{align}
where
\begin{align}
\rho(r_{\ast},\omega)=\frac{1}{2h}\sqrt{\bigg(\frac{d h(r_{\ast})}{dr_{\ast}}\bigg)^2+(\omega^2-V_{eff}-h^2(r_{\ast}))^2}.    
\end{align}
Above, $h(r_{\ast})$ represents a positive function satisfying $h(\infty)=h(-\infty)=w$
\begin{align}
\sigma(\omega)=\mathrm{sech}^2\bigg(\frac{1}{2\omega}\int_{r_H}^{\infty}V_{eff}(r_{\ast}) dr_{\ast} \bigg).   
\end{align}

As we can observe, the radial function $F(r)$ plays a pivotal role in determining the relationship between the GFs and the effective potential. The same procedure adopted in this section will be used again in the next section when we investigate the GF for the gravitino. Then, using the effective potential, one writes the transmission probability as follows:
\begin{align}
 \sigma(\omega)=\mathrm{sech}^2\bigg[\frac{1}{2\omega}\bigg(\int_{r_h}^{\infty}\bigg(\frac{dW}{dr_{\ast}}\bigg) dr_{\ast}+\int_{r_h}^{\infty}W^2 dr_{\ast}\bigg)\bigg] .  
\end{align}
We can separately analyze these two integrals. Taking into account $\lim_{r\rightarrow\infty}W=W(r_h)=0$, the first integral reduces to
\begin{align}
\int_{r_h}^{\infty}\bigg(\frac{dW}{dr_{\ast}}\bigg) dr_{\ast}=W(r\rightarrow\infty)-W(r_h)=0.    
\end{align}
On the other hand, the second integral takes the following form:
\begin{align}
\int_{r_h}^{\infty}\frac{1}{r^2}\frac{2\omega(\lambda^2+m^2r^2)^2}{2\omega(\lambda^2+m^2r^2)+m\lambda F}dr.    
\end{align}
This integral cannot be computed analytically. Thus, we utilize an
asymptotic series expansion method. After series
expansion, one writes the integrand as follows:
\begin{align}
\frac{1}{r^2}\frac{2\omega(\lambda^2+m^2r^2)^2}{2\omega(\lambda^2+m^2r^2)+m\lambda F}=\frac{\lambda ^2+m^2 r^2}{r^2 \left(1+\frac{\lambda  m F}{2 \omega  \left(\lambda ^2+m^2 r^2\right)}\right)}\nonumber\\\approx m^2+\frac{\lambda  m M}{r^3 \omega }+\frac{1}{r^2}\bigg(\lambda ^2-\frac{\lambda  m \left(\eta ^2 \kappa -1\right)}{2 (\gamma -1) \omega}\bigg)+\cdots.   
\end{align}

After integrating, we obtain the transmission probability as follows:
\begin{align}
\sigma(\omega)=\mathrm{sech}^2\bigg[-\frac{1}{2\omega}\left(m^2\, r_h-\frac{\lambda  m M}{2 \,r_h^2 \,\omega }+\frac{\lambda  m \left(\eta ^2 \kappa -1\right)}{2\, r_h\, \omega (\gamma -1)  }-\frac{\lambda ^2}{r_h}+\cdots\right)\bigg]\nonumber\\=\mathrm{sech}^2\bigg[\frac{1}{2\omega}\left(-\frac{2 (1-\gamma ) m^2 M}{1-\kappa \eta^2}-\frac{\lambda  m (1-\kappa \eta^2)^2}{8 M \omega\,(1-\gamma )^2  }+\frac{\lambda ^2 (1-\kappa \eta^2)}{2 M(1-\gamma ) }\right)\bigg].    
\end{align}

\begin{figure}[ht!]
\begin{center}
\hspace*{-3mm}
\begin{tabular}{ccc}
\includegraphics[scale=0.3]{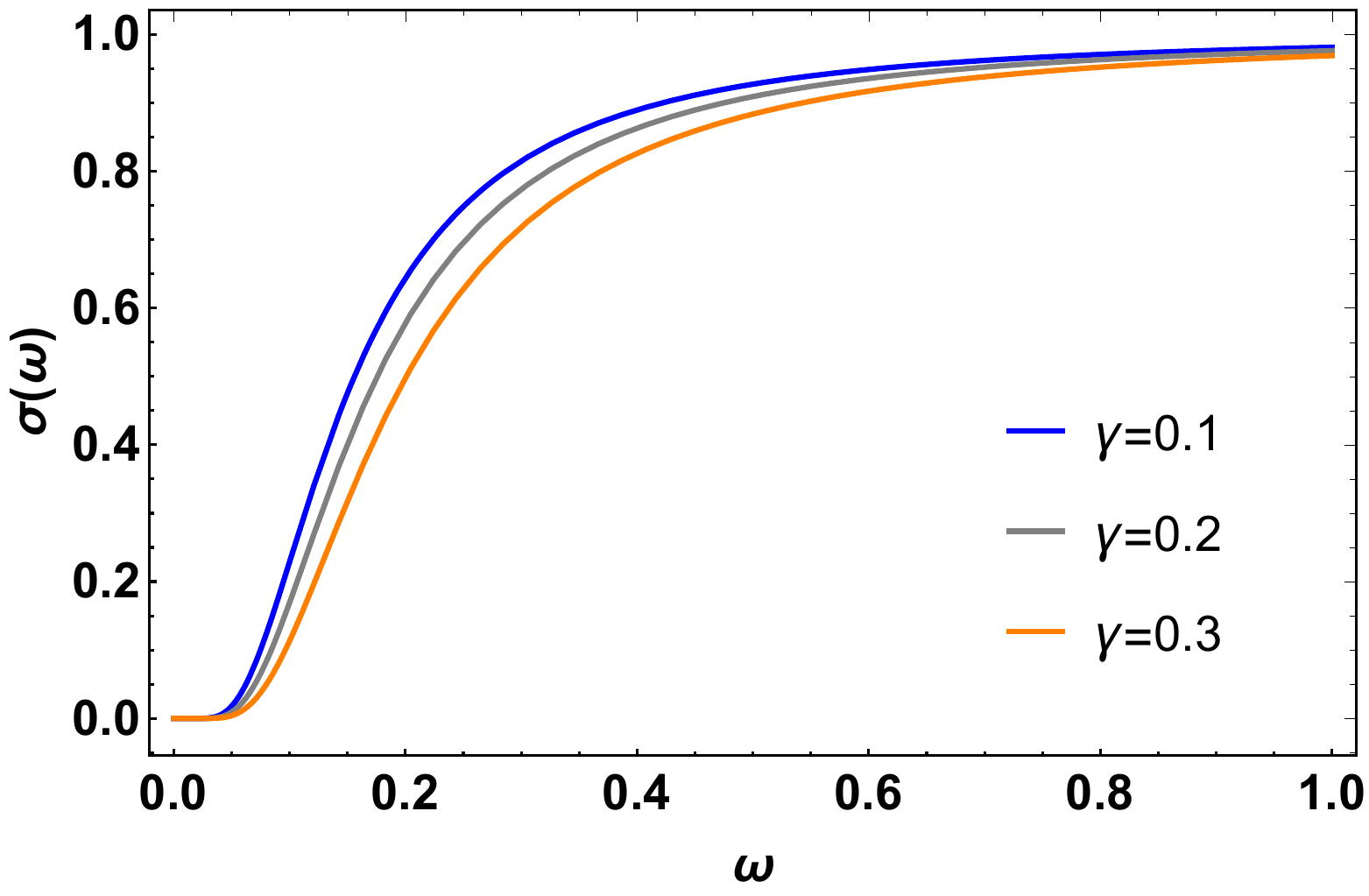} 
\includegraphics[scale=0.3]{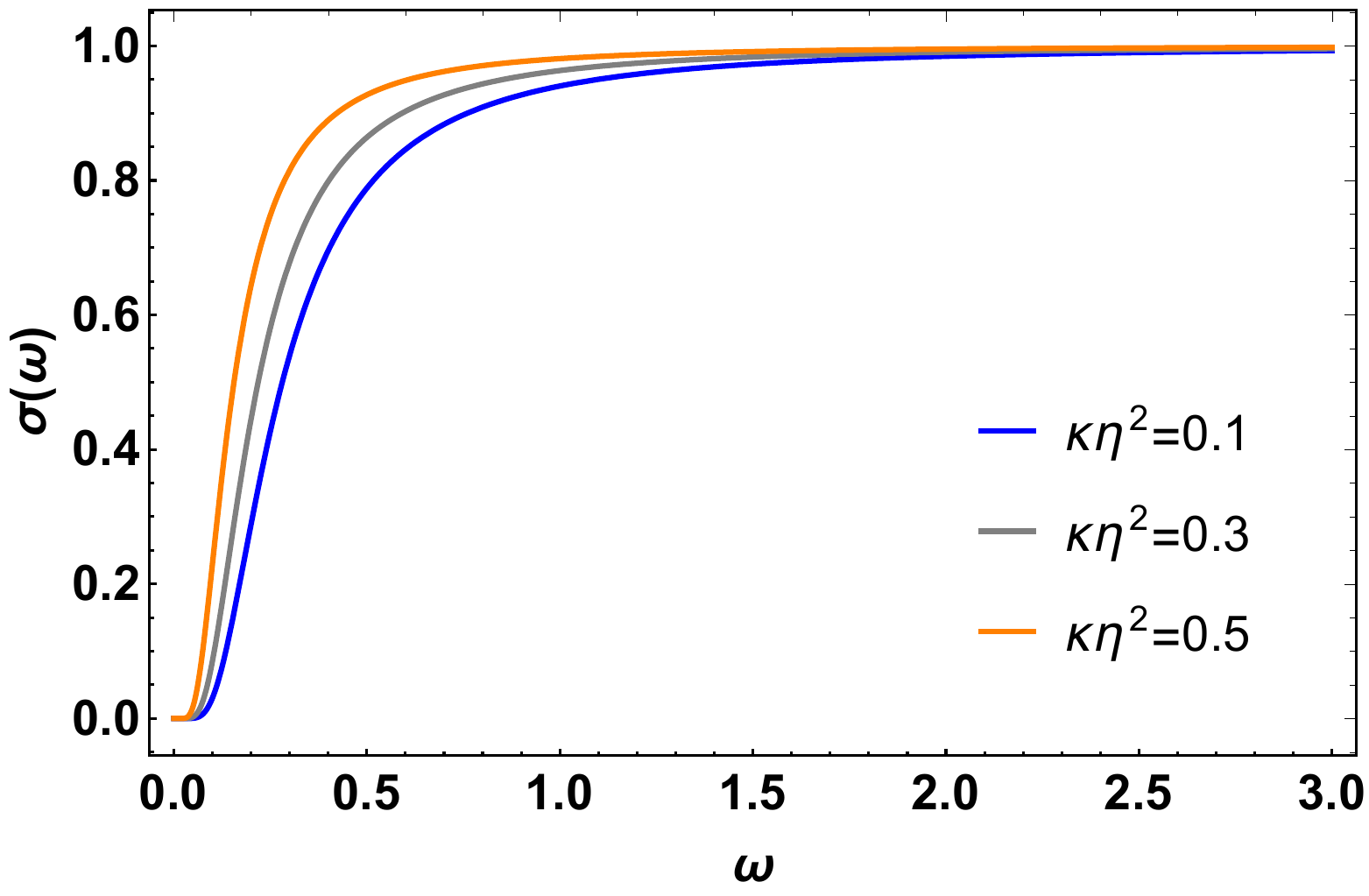}\\
(a)\hspace{8.5cm}(b)\\
\end{tabular}
\end{center}
\vspace{-0.5cm}
\caption{The representation of the greybody factors of massive spin 1/2 fermions as a function of $\omega$ by varying the parameters $\gamma$ for $\kappa\eta^2=0.5$ (a) and $\eta$ for $\gamma=0.1$ (b).
\label{fig1}}
\end{figure}

\begin{figure}[ht!]
\begin{center}
\hspace*{-3mm}
\begin{tabular}{ccc}
\includegraphics[scale=0.3]{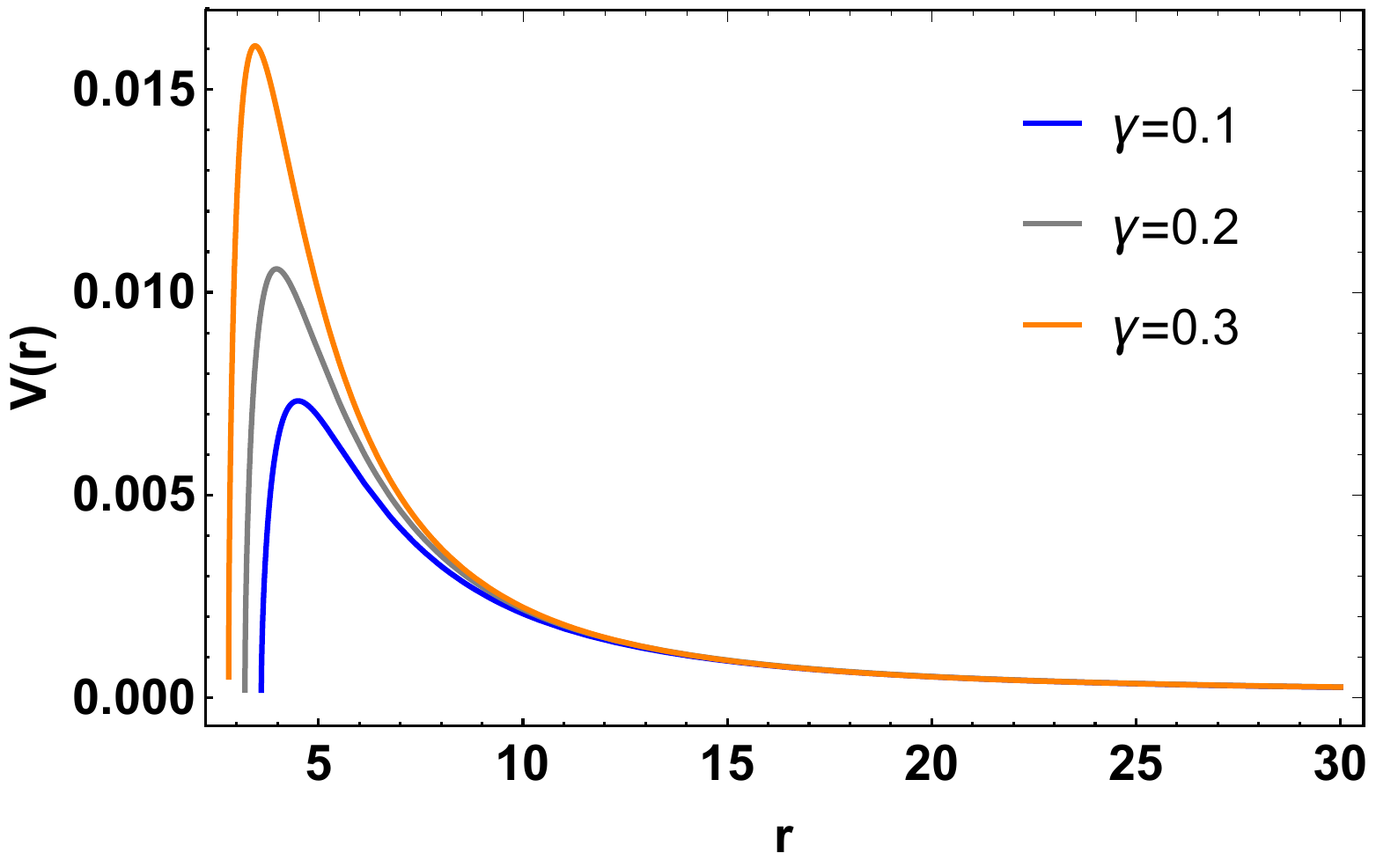} 
\includegraphics[scale=0.3]{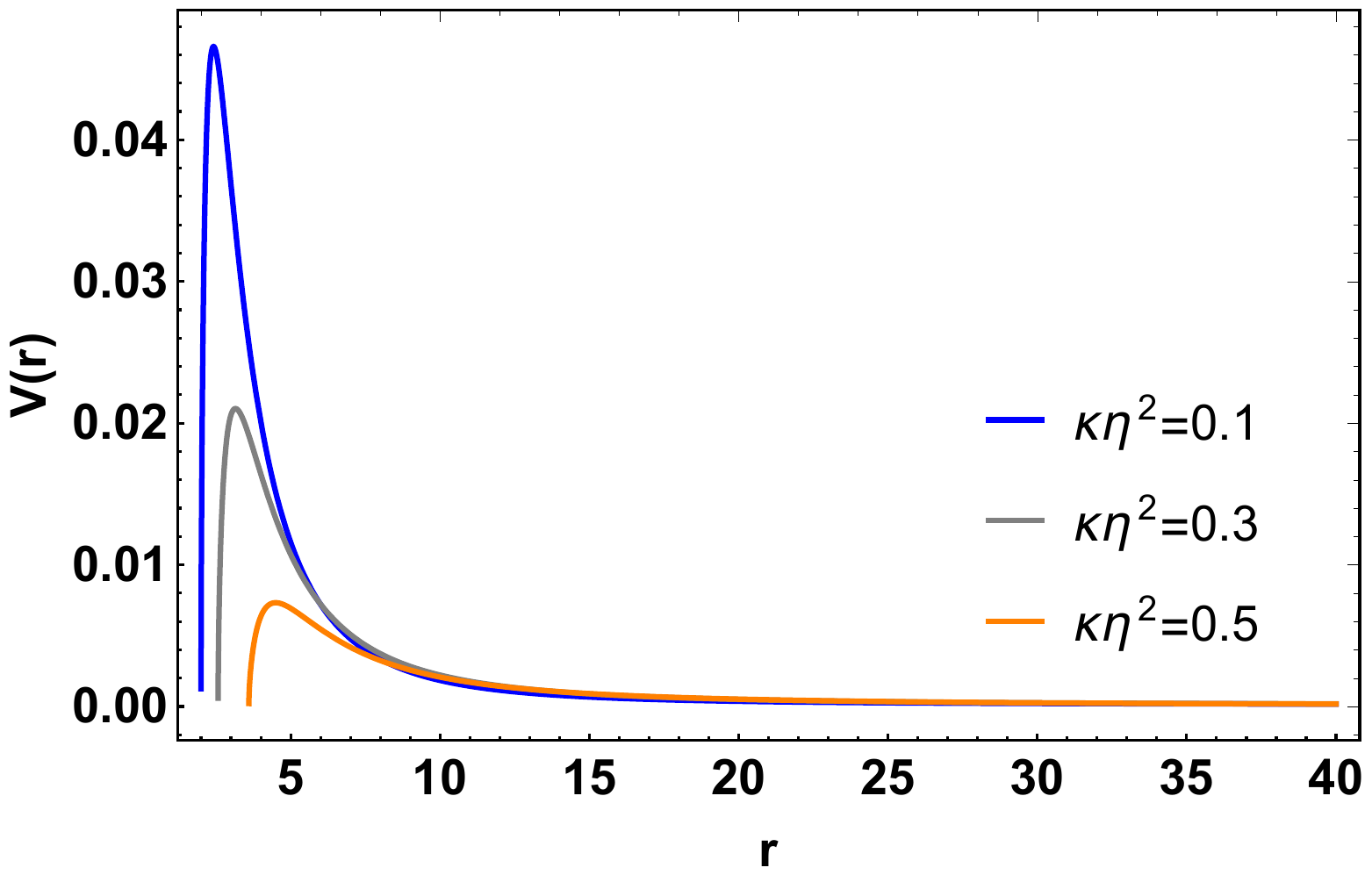}\\
(a)\hspace{8.5cm}(b)\\
\end{tabular}
\end{center}
\vspace{-0.5cm}
\caption{The representation of the effective potential of massive spin 1/2 fermions as a function of $r$ by varying the parameters $\gamma$ for $\kappa\eta^2=0.5$ (a) and $\eta$ for $\gamma=0.1$ (b).
\label{fig2}}
\end{figure}

The behavior of the GF factor of massive spin 1/2 fermions is depicted in Fig. (\ref{fig1}), where one notices the influence of both the LV parameter $\gamma$ and the global monopole charge $\eta$. At the lowest value of the frequency, there is no tunneling effect and the radiation is completely reflected, whereas one observes the tunneling as the frequency increases. In this context, the GF factor decreases as the LV parameter increases, indicating a lower probability of transmission Fig.(\ref{fig1})(a). On the other hand, the GF factor increases as the global monopole charge increases, indicating a higher probability of transmission Fig.(\ref{fig1})(b). The behavior of the effective potential shown in (\ref{fig2}) confirms our analysis. The higher the height of the potential barrier, the less radiation is able to reach infinity.

Now, let us focus on the massless case, where we use the potential (\ref{mlp}). Thus, we write the GF as follows
\begin{align}
\sigma(\omega)=\mathrm{sech}^2\bigg[\frac{\lambda}{2\omega}\int_{r_h}^{\infty}\frac{1}{r^2} \bigg(\lambda+\frac{r F^{\prime}}{2\sqrt{F}}-\sqrt{F}\bigg)dr\bigg].   
\end{align}
After some calculations, we arrive at
\begin{align}
\sigma(\omega)=\mathrm{sech}^2\bigg[\frac{1}{2\omega}\bigg(\frac{\lambda ^2 }{r_h}\bigg)\bigg]=\mathrm{sech}^2\bigg[\frac{1}{2\omega}\bigg(\frac{\lambda ^2 \left(1-\kappa\eta ^2\right)}{2 M(1-\gamma)}\bigg)\bigg].   
\end{align}

\begin{figure}[ht!]
\begin{center}
\hspace*{-3mm}
\begin{tabular}{ccc}
\includegraphics[scale=0.3]{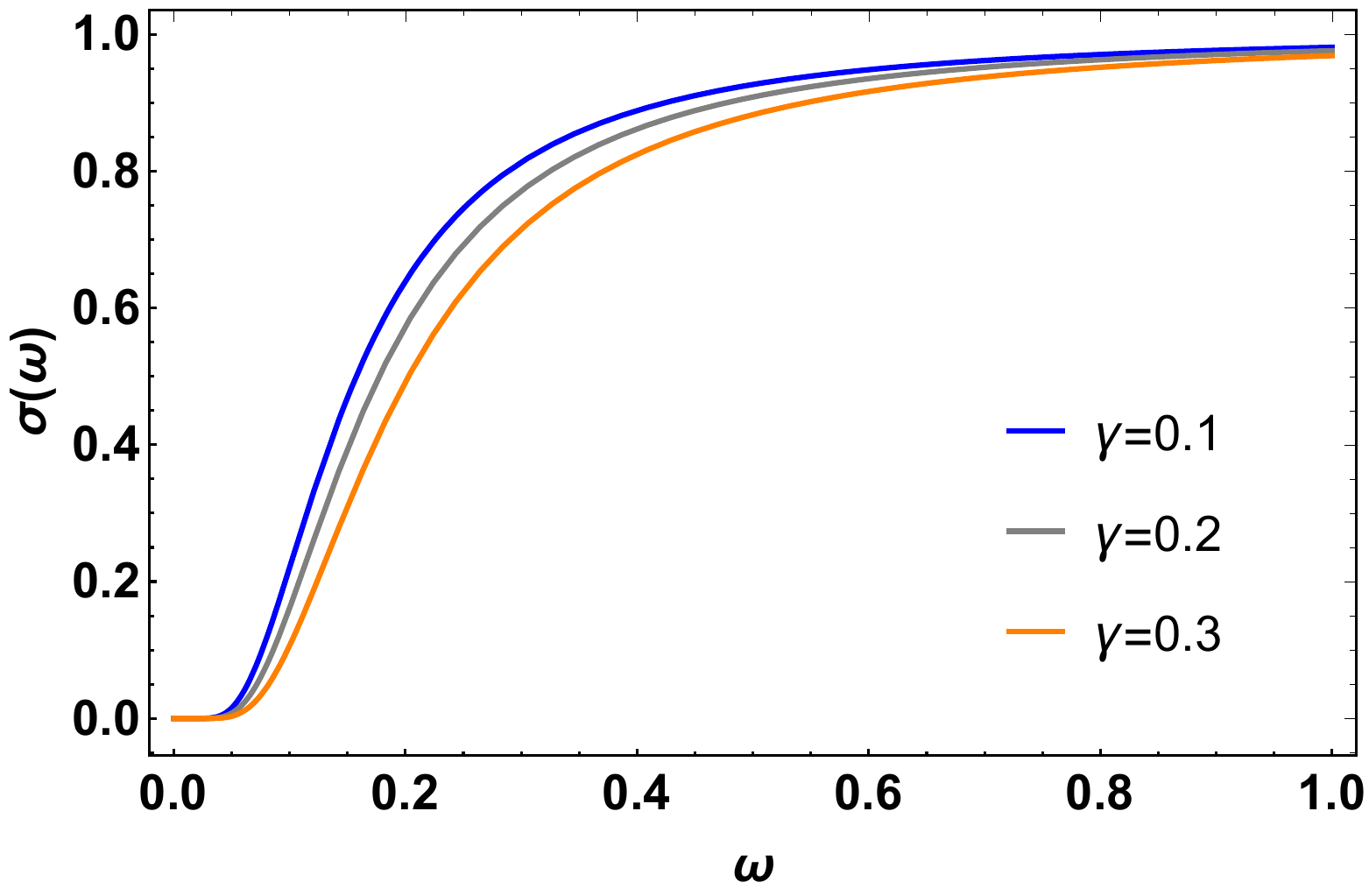} 
\includegraphics[scale=0.3]{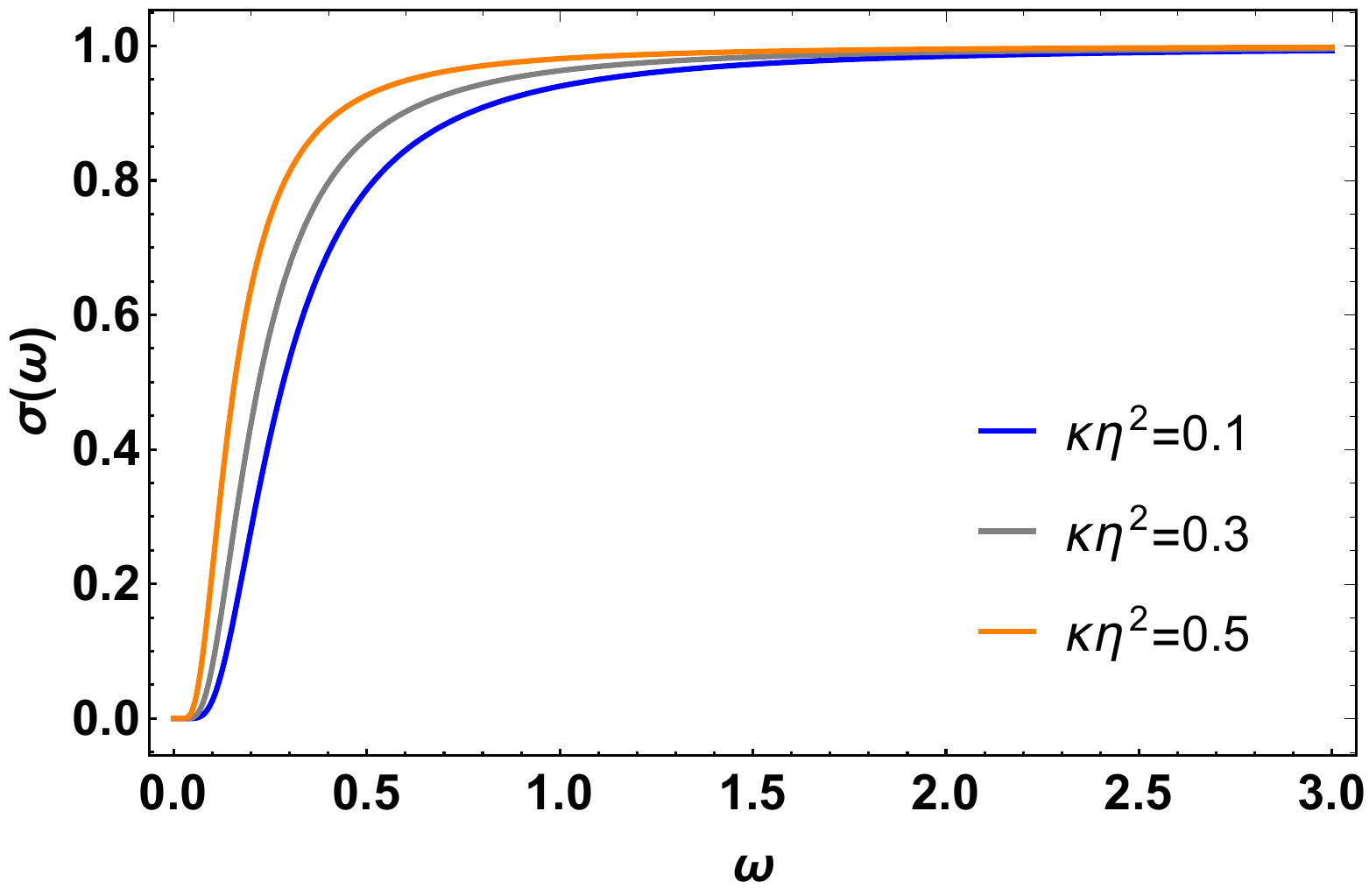}\\
(a)\hspace{8.5cm}(b)\\
\end{tabular}
\end{center}
\vspace{-0.5cm}
\caption{The representation of the greybody factors of massless spin 1/2 fermions as a function of $\omega$ by varying the parameters $\gamma$ for $\kappa\eta^2=0.5$ (a) and $\eta$ for $\gamma=0.1$ (b).
\label{fig3}}
\end{figure}

\begin{figure}[ht!]
\begin{center}
\hspace*{-3mm}
\begin{tabular}{ccc}
\includegraphics[scale=0.3]{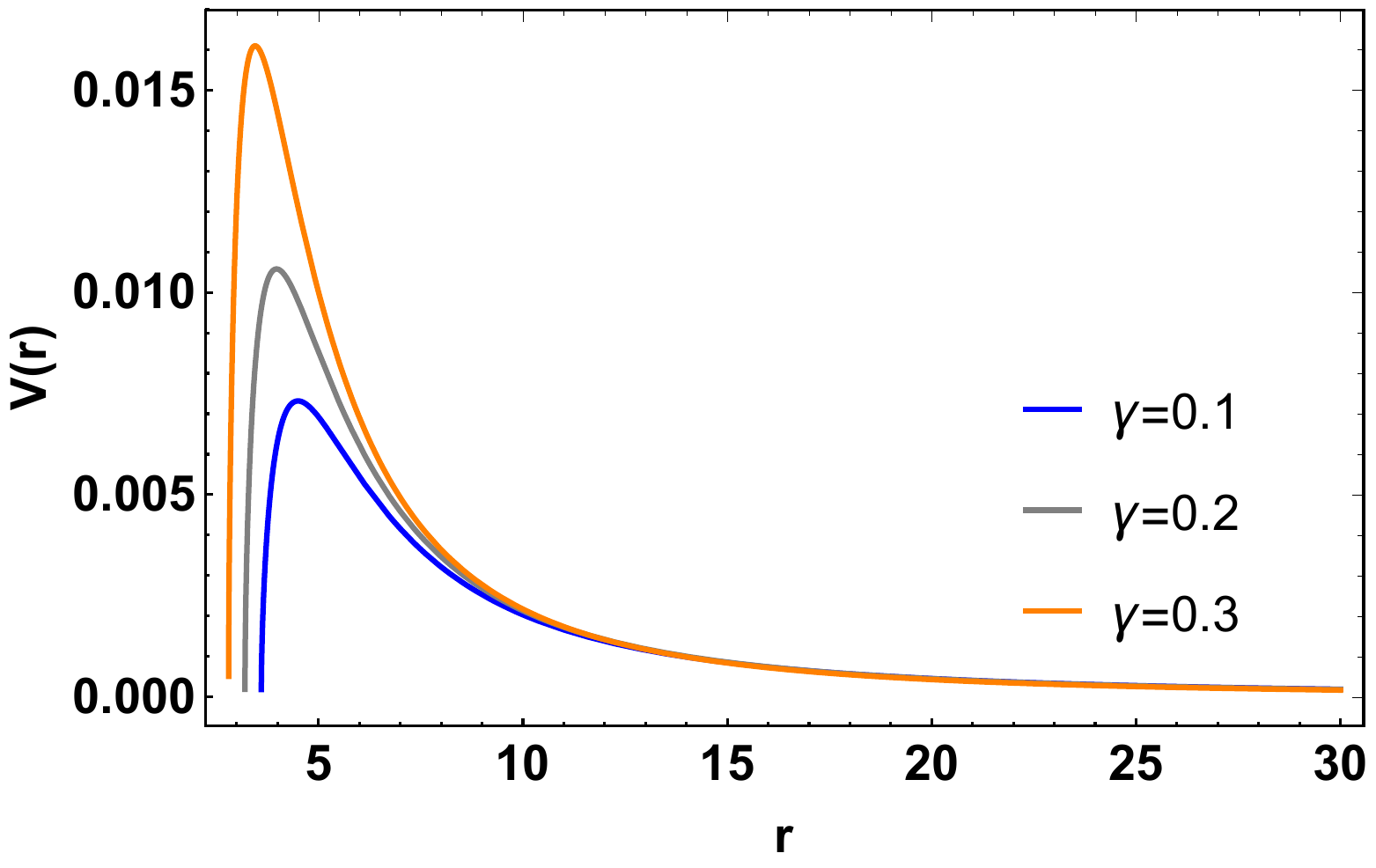} 
\includegraphics[scale=0.3]{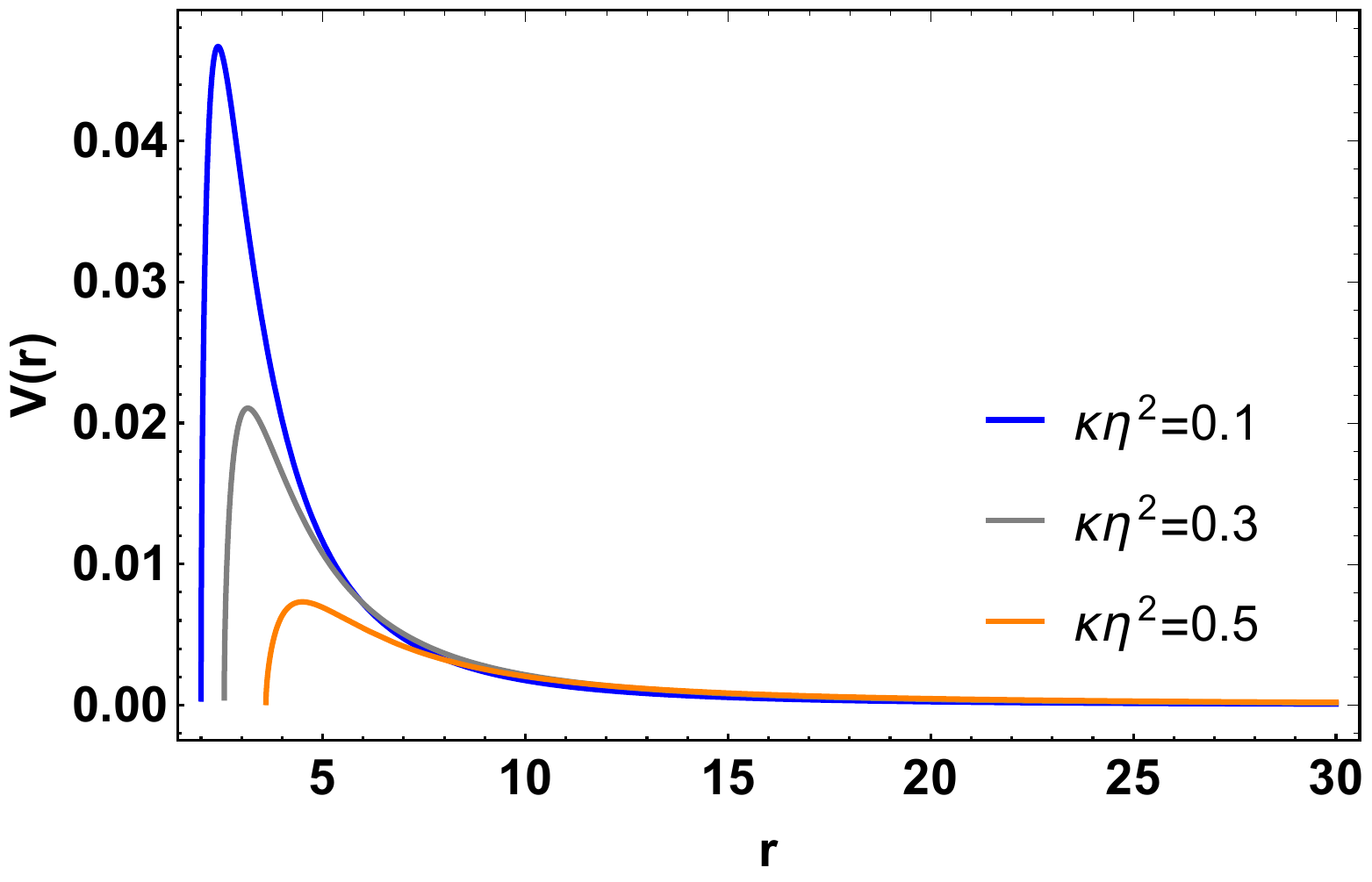}\\
(a)\hspace{8.5cm}(b)\\
\end{tabular}
\end{center}
\vspace{-0.5cm}
\caption{The representation of the effective potential of massless spin 1/2 fermions as a function of $r$ by varying the parameters $\gamma$ for $\kappa\eta^2=0.5$ (a) and $\eta$ for $\gamma=0.1$ (b).
\label{fig4}}
\end{figure}

We depict the GF factor of massless spin 1/2 fermions in Fig.(\ref{fig3}) and the effective potential in Fig.(\ref{fig4}). Like in the massive case, one observes a lower probability of transmission as the LV parameter increases Fig.(\ref{fig3})(a) and higher probability of transmission as the global monopole increases Fig.(\ref{fig3})(b). Besides, we note that there is a smaller variation of the GF factor in comparison with the massive case. In this conjecture, we note that the fermion mass also plays a role in the transmission of the particle, making such a process more difficult due to the interaction of the mass with curved spacetime.

\subsection{Spin 3/2 fermions}
Let us now focus on the Rarita–Schwinger equation, which represents the massless spin 3/2 field. The Rarita–Schwinger equation is given by \cite{Boonserm:2021owk, Al-Badawi:2023emj, Al-Badawi:2023xig}
\begin{align}
\gamma^{\mu\nu\alpha}\mathcal{D}_\nu\psi_\alpha=0,    
\end{align}
where $\psi_\alpha$ represents the spin-3/2 field, $\gamma^{\mu\nu\alpha}$ is the antisymmetric of Dirac gamma matrices and $\mathcal{D}_\nu$ is the supercovariant derivative, explicitly
\begin{align}
 \mathcal{D}_\nu=D_\nu+\frac{1}{4}\gamma_\alpha F^\alpha_\nu+\frac{i}{8}\gamma_{\nu\alpha\mu}F^{\alpha\mu}.   
\end{align}
Similarly to spin 1/2 fermions case, we obtain, after some calculations, a one-dimensional Schr\"{o}dinger-like wave equation given by
\begin{align}
\frac{d^2 H_{\pm}}{d r^2_{\ast}}+(\omega^2-\mathcal{V}_{\pm})H_{\pm}=0,   
\end{align}
where we have defined the effective potential $\mathcal{V}_{\pm}$ for spin 3/2 fermions as follows:
\begin{align}
\mathcal{V}_{\pm}=\pm\frac{d \mathcal{W}}{d r_{\ast}}+\mathcal{W}^2,    
\end{align}
where
\begin{align}
 \mathcal{W}=\frac{ g \,\sqrt{F}}{r}\bigg(\frac{g^2-1}{g^2-F}\bigg),   
\end{align}
where $g$ stands for the eigenvalue, where $g=j+1/2$, with $j=3/2, 5/2, 7/2,\ldots$. For this case, the tortoise coordinate reads
\begin{align}
    \frac{dr_{\ast}}{dr}=\frac{1}{F}.
\end{align}

With the effective potential for gravitino in hands, we are able to compute the transmission. To do so, we will follow the same steps as previously discussed.
\begin{align}
    \sigma(\omega)=\mathrm{sech}^2\bigg[\frac{g^2}{2\omega}\int_{r_h}^{+\infty}\frac{ 1}{r^2}\bigg(\frac{g^2-1}{g^2-F}\bigg)^2\,dr\bigg].
\end{align}
Upon integrating over $r$ and using the horizon radius, we explicitly arrive at
\begin{align}
    \sigma(\omega)=\mathrm{sech}^2\bigg[\frac{g^2}{2\omega}\bigg(\frac{(1-\gamma ) \left(1-g^2\right)^2 (1-\kappa\eta ^2)}{2 M \left(1-(1-\gamma) g^2-\kappa\eta ^2\right)^2}\bigg)\bigg].
\end{align}

\begin{figure}[ht!]
\begin{center}
\hspace*{-3mm}
\begin{tabular}{ccc}
\includegraphics[scale=0.3]{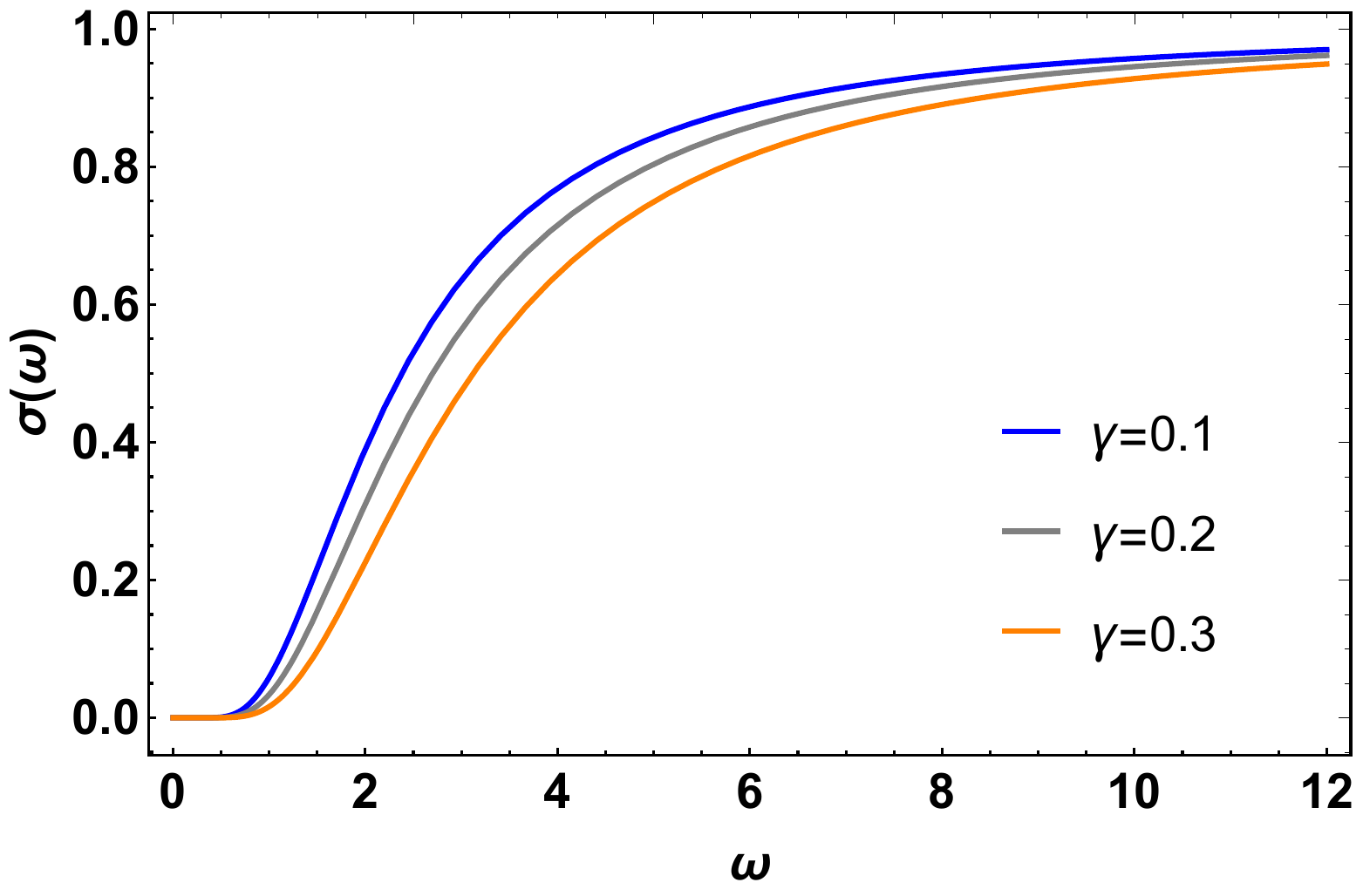} 
\includegraphics[scale=0.3]{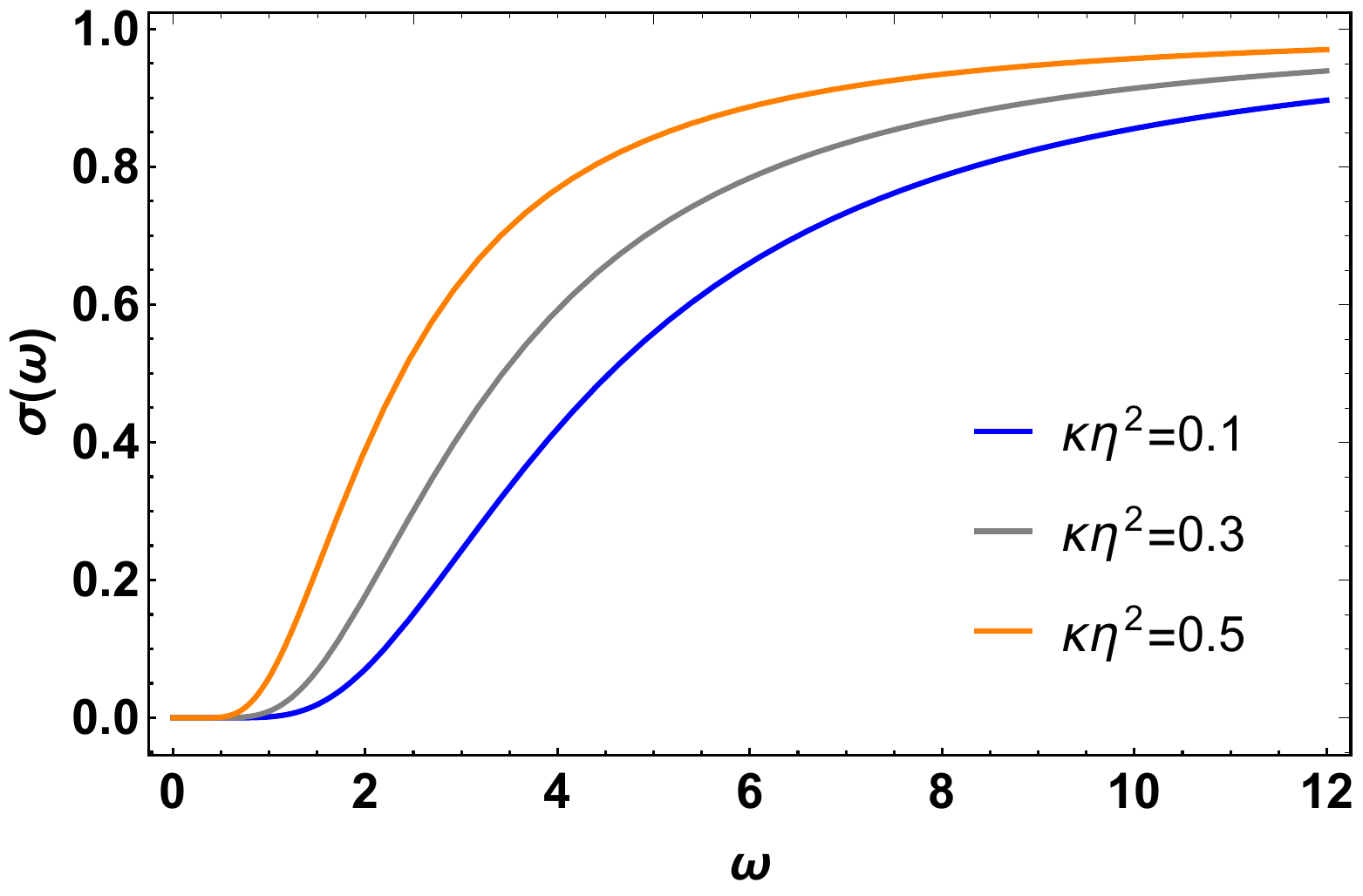}\\
(a)\hspace{8.5cm}(b)\\
\end{tabular}
\end{center}
\vspace{-0.5cm}
\caption{The representation of the greybody factors of spin 3/2 fermions (gravitino) as a function of $\omega$ by varying the parameters $\gamma$ for $\kappa\eta^2=0.5$ (a) and $\eta$ for $\gamma=0.1$ (b).
\label{fig5}}
\end{figure}

\begin{figure}[ht!]
\begin{center}
\hspace*{-3mm}
\begin{tabular}{ccc}
\includegraphics[scale=0.3]{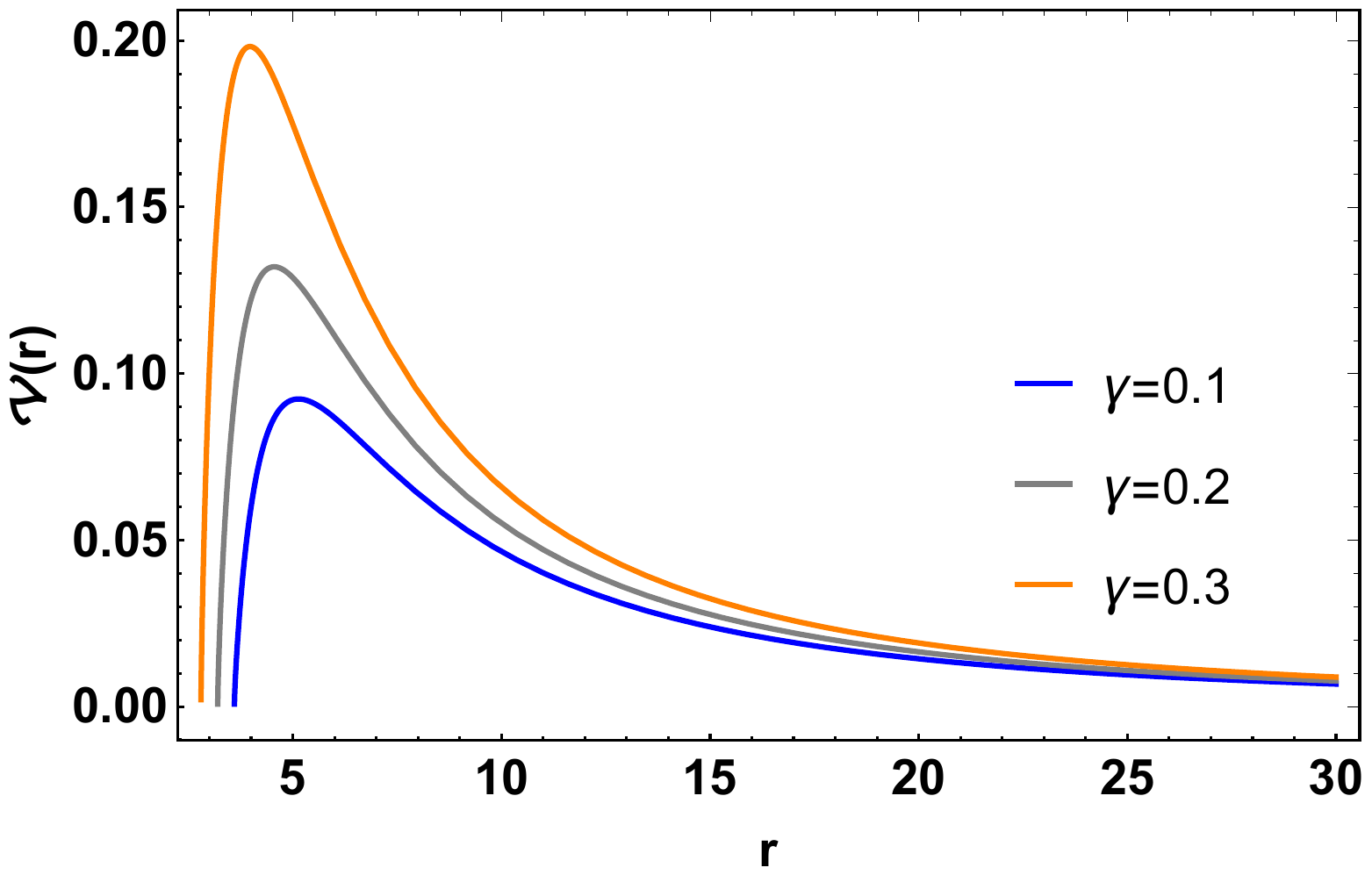} 
\includegraphics[scale=0.3]{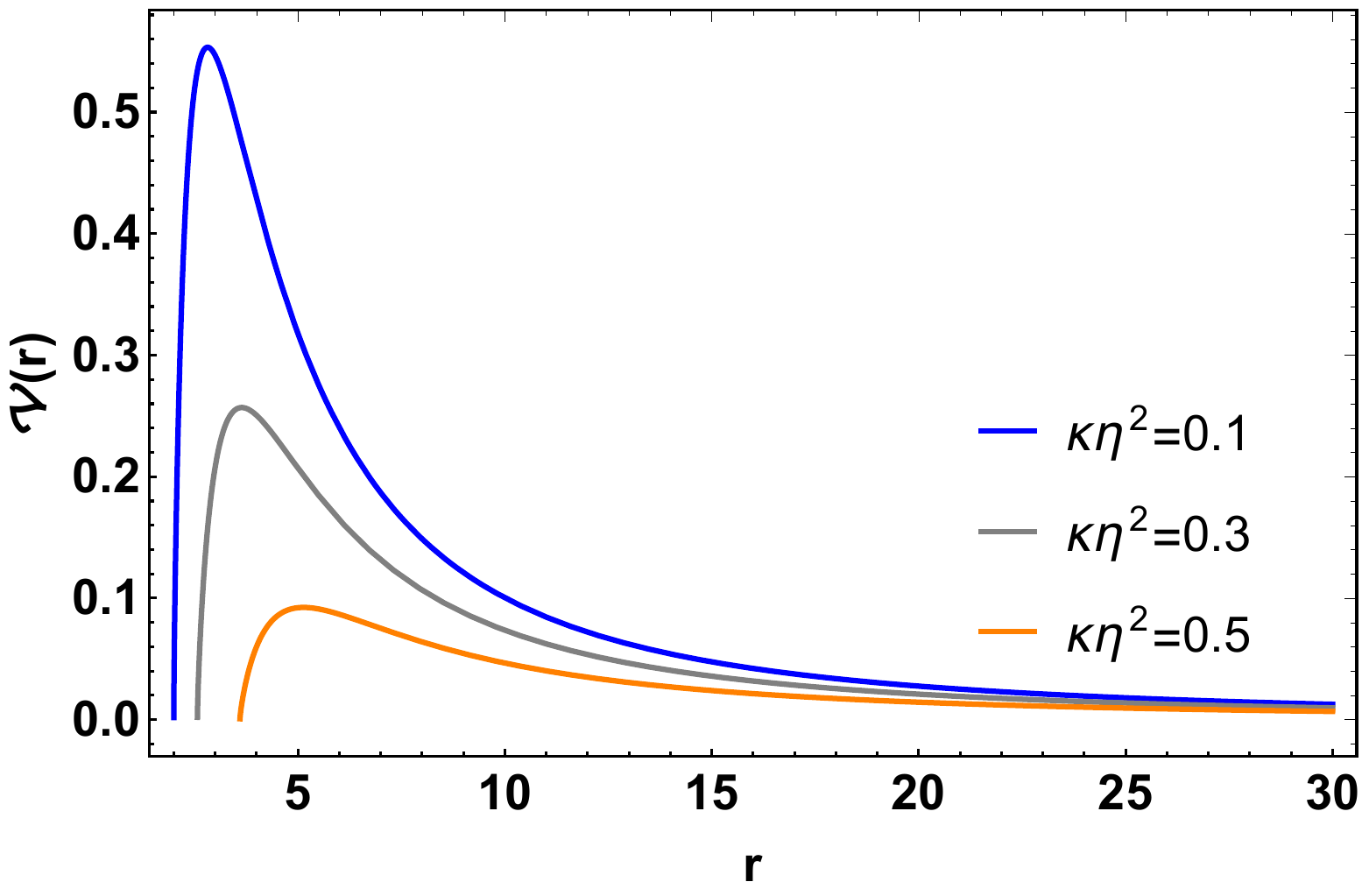}\\
(a)\hspace{8.5cm}(b)\\
\end{tabular}
\end{center}
\vspace{-0.5cm}
\caption{The representation of the effective potential of spin 3/2 fermions (gravitino) as a function of $r$ by varying the parameters $\gamma$ for $\kappa\eta^2=0.5$ (a) and $\eta$ for $\gamma=0.1$ (b).
\label{fig6}}
\end{figure}

We illustrate the GF factor of gravitino in Fig. (\ref{fig5}) and its effective potential in Fig. (\ref{fig6}). In this case, one notices a similar behavior of the GF compared to massless 1/2 fermions. Less radiation is able to reach an observer at an infinite distance when the value of the LV parameter is increased Fig.(\ref{fig5})(a) and more radiation is able to reach an observer at an infinite distance when the value of the global monopole charge is increased, see Fig.(\ref{fig5})(b).

\section{Gravitational lensing in the strong field limit}\label{s4}

The phenomenon of light bending while it passes through a gravitational field is known to be one of the most important predictions of general relativity (GR). In this section, we are interested in investigating the various aspects of gravitational lensing in the strong field limit for the space-time described by our metric (\ref{metric}). In this conjecture, we will adopt the methodology developed by Bozza and Tsukamoto \cite{Bozza:2002zj, Tsukamoto:2016jzh}.

To reach our goal, we will initially analyze the geodesics associated with this spacetime. Because of the spherical symmetry, one can focus solely on the radial geodesics. Further, we shall show possible situations in which the spacetime is geodesically complete. Initially, we will consider the following Lagrangian
\begin{align}
  \mathcal{L}=\frac{1}{2}g_{\alpha\beta}\frac{d x^\alpha}{d\tau}\frac{d x^\beta}{d\tau}. 
\end{align}

One can consider the light propagation in the equatorial plane $\theta = \pi/2$, obtaining the following expression
\begin{align}\label{eqg}
\mathcal{L}=-F(r)\Dot{t}^2+\frac{1}{F(r)}\Dot{r}^2+r^2\Dot{\phi}^2,
\end{align}
where the dot represents derivative with respect to an affine parameter denoted by $\lambda$. Following general relativity, $\mathcal{L}$ takes the values $-1/2$ (time-like geodesics), $0$ (null geodesics), $1/2$ (space-like geodesics). It is easy to observe that the metric (\ref{metric}) has only two Killing vector fields, namely $\partial/\partial t$ and $\partial/\partial \phi$. The Killing vector $\partial/\partial t$ is associated with energy conservation, and the Killing vector $\partial/\partial \phi$ is associated with angular momentum conservation. From (\ref{eqg}), we obtain both the conserved energy $E$ and angular momentum $L$ for a particle along the geodesics, explicitly
\begin{align}
\label{Ec} E=-\frac{\partial\mathcal{L}}{\partial\Dot{t}}=F(r)\Dot{t},\\
\label{Lc} L=\frac{\partial\mathcal{L}}{\partial\Dot{\phi}}=r^2\Dot{\phi}.
\end{align}

Now, let us study the null geodesics, taking $\mathcal{L}=0$. Thereby, we obtain the equation for each photon as follows
\begin{align}\label{me0}
\frac{1}{2}\left(\frac{d r}{d\tau}\right)^2+\mathcal{U}_{g}=\mathcal{E},  
\end{align}
where we have defined the gravitational potential $\mathcal{U}_{g}$ and the energy $\mathcal{E}$ for photon written below
\begin{align}
\mathcal{U}_{g}=F(r)\left(\frac{L^2}{2r^2}\right),    
\end{align}
and
\begin{align}
\mathcal{E}=\frac{E^2}{2}.    
\end{align}
Let us now analyze the unstable orbits satisfying the conditions $\frac{d \mathcal{U}_{g}}{dr}=0$ and $\frac{d^2 \mathcal{U}_{g}}{dr^2}<0$. For the solution (\ref{sol1}), the gravitational potential is given by
\begin{align}\label{pot}
\mathcal{U}_{g}=\left(\frac{1+\eta^2}{1-\gamma}-\frac{2M}{r}\right)\left(\frac{L^2}{2r^2}\right). 
\end{align}
Then, we obtain from the above potential (\ref{pot}) the critical radius, namely
\begin{align}\label{cr1}
  r_c=\frac{3M(1-\gamma)}{1-\kappa\eta^2}.  
\end{align}

To apply the methodology developed in \cite{Tsukamoto:2016jzh}, we take into account that our line element (\ref{metric}) perfectly matches the form assumed in \cite{Tsukamoto:2016jzh} given by
\begin{equation}
ds^2=-A(r)dt^2+B(r)dr^2+C(r)(d\theta^2+\sin^2\theta d\phi^2),
\end{equation}
where $A(r)=F(r)$, $B(r)=1/F(r)$ and $C(r)=r^2$. We consider a photon starting from infinity, approaching a black hole with a given parameter of impact $b$,  deflected at a closest approach $r_0$ and then returning to infinity. One defines the impact parameter through the relation
\begin{align}
 b_c=\sqrt{\frac{C_0}{A_0}}.   
\end{align}
Above, the subscript $O$ indicates that the function is evaluated at $r_0$. The deflection of light  expressed in terms of the closest approach $r_0$ as
\begin{align}
    \alpha(r_0)=I(r_0)-\pi,
\end{align}
where
\begin{align}
I(r_0)=2\int_{r_0}^{\infty}\frac{dr}{\sqrt{\frac{R(r)C(r)}{B(r)}}}.    
\end{align}

In the weak field limit, that is, if we take into account only terms up to the first order in $M$ and $\gamma$, the angular deflection reads
\begin{align}
\alpha&=\pi\sqrt{\frac{1-\gamma}{1-\kappa\eta^2}}+\frac{4M}{\beta}\bigg(\frac{1-\gamma}{1-\kappa\eta^2}\bigg)^2-\pi + \mathcal{O}(M^2)\nonumber\\ &\approx \pi\bigg[\frac{1-\frac{\gamma}{2}}{\sqrt{1-\kappa\eta^2}}-1\bigg]+\frac{4M}{\beta} \frac{1-2\gamma}{(1-\kappa\eta^2)^2}.  
\end{align}

It is straightforward to observe that the first term refers to angular deflection for $M=0$, leading to a non-zero light deflection. At the same time, the second term indicates the angular deflection for the black hole with mass $M$. Thus, we can see that both terms bring LV contributions in this limit. However, when $r_0$ is strongly different from $b$, this limit is no longer valid. The next step is to consider the strong-field limit. The approach to be adopted works as follows: the lower $r_0$, the greater the deflection. Then, when $r_0$ coincides with the radius of the photon sphere $r_m$ the deflection angle diverges. In Ref \cite{Bozza:2002zj}, it was shown that this divergence is logarithmic, and also an algorithm for the calculation of the angular deflection was presented as well. Posteriorly, this approach was improved by Tsukamoto \cite{Tsukamoto:2016jzh}. Below, we will discuss this approach and utilize it within our calculations. As a starting point, we introduce the variable
\begin{align}
    z=1-\frac{r_0}{r},
\end{align}
so that the integral takes the form
\begin{align}
I(r_0)=\int_{0}^{1}f(z,r_0)dz,    
\end{align}
where
\begin{align}
f(z,r_0)=\frac{2r_0}{(1-z)^2\sqrt{\frac{R(r)C(r)}{B(r)}}}.  
\end{align}
In order to compute the integral $I(r_0)$, we split it into two parts: the divergent one and the regular one, namely
\begin{align}
I_D(r_0)=\int_{0}^{1}f_D(z,r_0)dz    
\end{align}
and
\begin{align}
I_R(r_0)=\int_{0}^{1}f_R(z,r_0)dz,    
\end{align}
where $f_R(z,r_0)=f(z,r_0)-f_D(z,r_0)$. In addition, $f_D(z,r_0)=\frac{2r_0}{\sqrt{c_1 z + c_2 z^2}}$. The deflection angle of the light in the strong limit reads
\begin{align}
 \alpha=-\overline{a}\,\mathrm{log}\left(\frac{b}{b_c}-1\right)+\overline{b}+\mathcal{O}[(b-b_c)\,\mathrm{log}(b-b_c)],  
 \label{angle}
\end{align}
where
\begin{align}
\overline{a}=\sqrt{\frac{2B_m A_m}{C^{\prime\prime}_m A_m-C_m A^{\prime\prime}_m}},
\end{align}
and
\begin{align}
\overline{b}=\overline{a}\,\mathrm{log}\left[r^2\left(\frac{C^{\prime\prime}_m}{C_m}-\frac{A^{\prime\prime}_m}{A_m}\right)\right]+I_R(r_m)-\pi.    
\end{align}

Once we have presented some key ideas of the methodology developed by Bozza and Tsukamoto \cite{Bozza:2002zj, Tsukamoto:2016jzh}, let us study the deflection angle of the light for the geometry described by the metric (\ref{sol1}). Firstly, we write the critical parameter $b_c$ as follows
\begin{align}
 b_c=3\sqrt{3}M\left(\frac{1-\gamma}{1-\kappa\eta^2}\right)^{\frac{3}{2}}.   
\end{align}
We need to calculate the radius $r_m$. To do so, we will follow the way presented in Ref. \cite{Tsukamoto:2016jzh}, where it is said that the radius of the photon sphere for a static spherically symmetric spacetime is determined by solving the following equation:
\begin{align}
    \bigg[\frac{C^{\prime}(r)}{C(r)}-\frac{A^{\prime}(r)}{A(r)}\bigg]\bigg\vert_{r=r_m}=0.
\end{align}
Therefore, using the metric (\ref{sol1}), we arrive at $r_m=r_c=\frac{3M(1-\gamma)}{1-\kappa\eta^2}$. This means that the radius $r_m$ is equal to the critical radius. Then, one expresses $\overline{a}$ and $\overline{b}$ as
\begin{align}
\overline{a}=\sqrt{\frac{1-\gamma}{1-\kappa\eta^2}}   
\label{bara}
\end{align}
and
\begin{align}
\overline{b}=\sqrt{\frac{1-\gamma}{1-\kappa\eta^2}}\,\mathrm{log}\,6 + I_R(r_m)-\pi. 
\label{barb}
\end{align}
With these results in hands, we are able to evaluate $I_R(r_m)$, thereby obtaining
\begin{align}
I_R(r_m) &= \sqrt{\frac{1-\gamma}{1-\kappa\eta^2}}\int_0^1\left(\frac{2}{z\sqrt{1-\frac{2z}{3}}}-\frac{1}{z}\right)dz \nonumber\\ &= 2\sqrt{\frac{1-\gamma}{1-\kappa\eta^2}}\, \mathrm{log}[6(2-\sqrt{3})]. 
\end{align}
Thus, we write $\overline{b}$ explicitly as follows
\begin{align}
\overline{b}= \sqrt{\frac{1-\gamma}{1-\kappa\eta^2}}\, \mathrm{log}[216(7-4\sqrt{3})]-\pi.   
\end{align}
Finally, we can write down the deflection angle:
\begin{align}
  a=\sqrt{\frac{1-\gamma}{1-\kappa\eta^2}}\,\left\{ \mathrm{log}\left[\frac{1}{3\sqrt{3}M}\left(\frac{1-\kappa\eta^2}{1-\gamma}\right)^{\frac{3}{2}}-1\right]+\mathrm{log}[216(7-4\sqrt{3})]-\pi\right\}.  
\end{align}
These results will be used in the next section.

\section{Lens equations and observables}\label{s5}

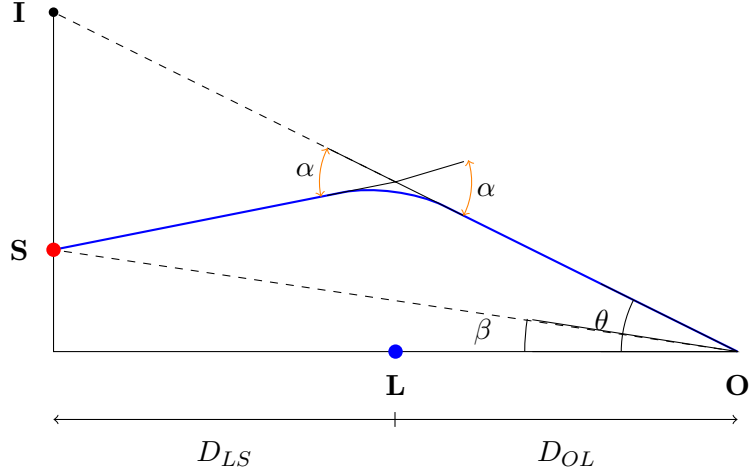
\begin{figure}
  \centering
  \begin{tikzpicture}[scale=0.9]
    \node (I)    at ( 5,-0.5)   {{\bf L}};
    \node (II)   at (10,-0.5)   {{\bf O}};
    \node (II)   at (-0.5,1.5)  {{\bf S}};
    \node (II)   at (-0.5,5)    {{\bf I}};
    \node        at ( 2.5,-1.5) {$D_{LS}$};
    \node        at ( 7.5,-1.5) {$D_{OL}$};

    \draw (10,0)--(0,0)--(0,5);
    \draw [thick,rounded corners=20pt, blue] (0,1.5)--(5,2.5)--(10,0); 
    \draw [dashed] (5,2.5)--(0,5);
    \draw [dashed] (10,0)--(0,1.5);
    \fill[blue] (5,0) circle (3pt);
    \fill[red] (0,1.5) circle (3pt);
    \fill[black] (0,5) circle (2pt);

    \draw
      (6,2) coordinate (a)
      -- (5,2.5) coordinate (b)
      -- (6,2.8) coordinate (c)
      pic["$\alpha$", draw=orange, <->, angle eccentricity=1.2, angle radius=1cm]
      {angle=a--b--c};

    \draw
      (4,2.3) coordinate (d)
      -- (5,2.5) coordinate (e)
      -- (4,3) coordinate (f)
      pic["$\alpha$", draw=orange, <->, angle eccentricity=1.2, angle radius=1cm]
      {angle=f--e--d};

    \draw
      (8,0) coordinate (g)
      -- (10,0) coordinate (h)
      -- (8,1) coordinate (i)
      pic["$\theta$", draw=black, angle eccentricity=1.2, angle radius=1.53cm]
      {angle=i--h--g};

    \draw
      (7,0) coordinate (g1)
      -- (10,0) coordinate (h1)
      -- (7,0.47) coordinate (i1)
      pic["$\beta$", draw=black, angle eccentricity=1.2, angle radius=2.8cm]
      {angle=i1--h1--g1};

    \draw[<->,
      decoration={markings, mark= at position 0.5 with {\arrow{|}},},
      postaction={decorate}
    ] (0,-1) -- (10,-1);
  \end{tikzpicture}
  \caption{In this lens diagram, $D_{OL}$ is the distance from the observer $O$ to the lens $L$, and $D_{LS}$ is the distance from the lens to the source projected onto the optical axis.}
  \label{LensingFig}
\end{figure}

Once we have obtained the angular deflection of light, the next step is to examine gravitational lenses. Below, let us briefly review the lens equation in the strong field limit. We begin by considering a light source located at $S$, emitting a light ray that is deflected towards the observer $O$ due to the presence of the compact object $L$. On the other hand, one denotes the angular position of the source by $\beta$, while the angular position of the image observed by $O$ is represented by $\theta$, and the angular deflection of light is $\alpha$, see Fig. \ref{LensingFig}. Furthermore, $D_{LS}$ denotes the distance between the source $S$ and the lens $L$, $D_{OL}$ is the distance between the observer $O$ and the lens $L$, and $D_{OL}=D_{OL} + D_{LS}$ is the total distance between the observer and the source. These distances are measured relative to the optical axis, which is the straight line passing through $D_{OL}$. Following the approach adopted in \cite{Tsukamoto:2016jzh}, we assume that the source $S$ is almost perfectly aligned with the lens $L$. Under this assumption, the relationship between the angular position of the source, the position of the image observed by the observer, and the angular deflection is given by:
\begin{align}
  \beta=\theta-\frac{D_{LS}}{D_{OS}}\Delta a_n,  
\end{align}
where $\Delta a_n$ is the deflection angle with subtracting all the loops performed by the photons before moving towards the observer, that is, $\Delta a_n = a-2n\pi$. Thus, the deflection angle reads
\begin{align}
    a(\theta)=-\overline{a}\,\mathrm{ln}\bigg(\frac{\theta D_{OL}}{b_c}-1\bigg)+\overline{b},
\end{align}
with
\begin{align}
\theta^{0}_{n}=\frac{b_c}{D_{OL}}(1+e_n),\   \  e_n=e^{\overline{b}-2n\pi},    
\end{align}
and
\begin{align}
 \Delta a_n=\frac{\partial a}{\partial\theta}\bigg\vert_{\theta=\theta^0} (\theta-\theta^{0}_{n}).   
\end{align}
As a result, we can obtain the position of the $n^{th}$ relativistic image
\begin{align}
    \theta_n=\theta^{0}_{n}+\frac{b_c\, e_n}{\overline{a}}\frac{D_{OS}}{D_{OL} D_{LS}}(\beta-\theta^{0}_{n}).
\end{align}
The other important observable is the magnification $\mu_n$, which is explicitly given by
\begin{align}
    \mu_n=\frac{e_n (1+e_n)}{\overline{a}\,\beta}\frac{D_{OS}}{D_{LS}}\left(\frac{b_c}{D_{OL}}\right)^2.
\end{align}

We point out that we express the positions of the relativistic images and also the magnification in terms of the coefficients of expansion. With these coefficients in our hands, we can compare our results with the experimental data. The inverse problem consists in finding the coefficients of the expansion in the strong field limit from the positions and the flow, thus discovering the nature of the object responsible for the lens. We write observables explicitly below:
\begin{align}
\theta_{\infty}&=\frac{b_c}{D_{OL}},\\
s&=\theta_{1}-\theta_{\infty}=\theta_{\infty} e^{\frac{\overline{b}-2\pi}{\overline{a}}},\\
\tilde{r}&=e^{\frac{2\pi}{\overline{a}}}.
\end{align}
Above, $s$ represents the angular separation, while $\tilde{r}=\dfrac{\mu_1}{\sum_{n=2}^{\infty}\mu_n}$ is the relationship between the flux of the first image and the flux of all higher-order images ($n>2$). In the next section, we will use the analysis of the observables to study the effects of the LV parameter and the global monopole charge in the LV black hole shadows.

 \section{LV black hole shadows analysis}\label{shadows}
In this section, we will investigate the shadows cast by the LV black hole solution sourced by a global monopole in terms of the LV parameter, $\gamma$, and the global monopole charge, $\eta$. Typically, there is a region around the black hole, called the photon sphere -- with critical radius $r_c$, in which photons approaching the black hole from infinity are forced to follow unstable circular geodesic orbits. The projection of these orbits casts a dark region on the observer’s sky, known as a black hole shadow. 

As it was primarily discussed in the previous paper \cite{Belchior:2025xam}, the authors obtained the shadow radius for the LV black hole in the presence of a global monopole \eqref{sol1} for an observer located at infinity and also imposed bounds on these parameters by comparing with observational data from the shadow radius of Sagittarius A$^{*}$ (Sgr A$^{*}$). The expression for the shadow radius is given by
\begin{equation}
    r_s=\frac{r_c}{\sqrt{f(r_c)}}=\frac{3\sqrt{3}M(1-\gamma)}{1-\kappa\eta^2},
\end{equation}
from which one recovers the shadow radius of the Schwarzschild black hole when $\gamma=\eta=0$, as expected. In this regard, our aim here is to go beyond the simple analysis of the shadow radius; we will image the LV black hole shadows when illuminated by a thin accretion disk.

The previous equation tells us about the influence of the LV parameter and the global monopole charge on the shadow radius of the LV black hole. The departure from the standard metric (Schwarzschild black hole) is more evident as $\gamma$ and $\eta$ become large. We will here explore the optical appearance of the LV black hole shadows using an optically thin ring/shell model, this analysis was not considered in \cite{Belchior:2025xam}.

\begin{figure}[ht!]
  \centering
\begin{subfigure}{0.33\textwidth}
    \centering
    \includegraphics[width=\linewidth]{\detokenize{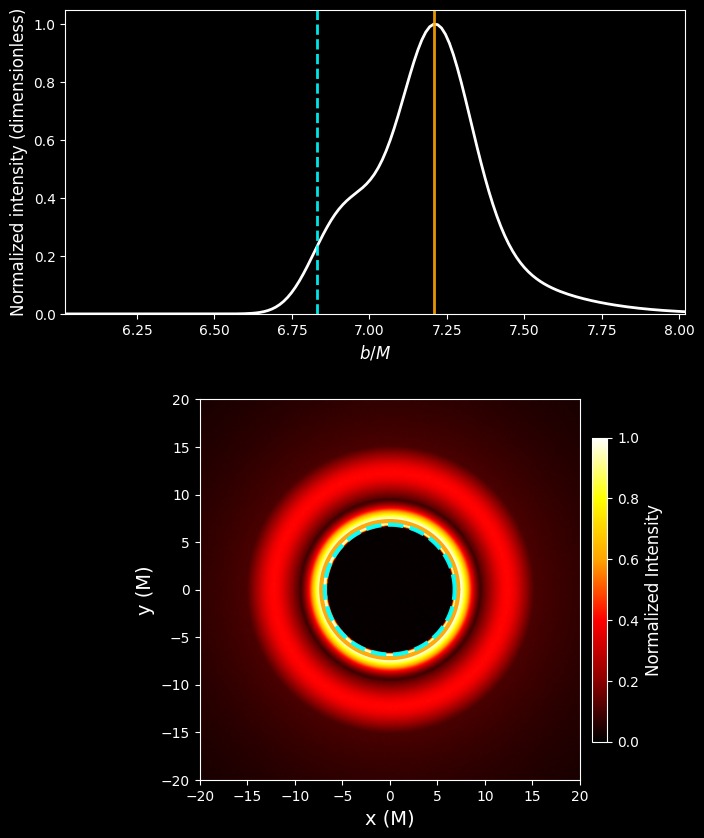}}
    \caption{}
  \end{subfigure}\hfill
  \begin{subfigure}{0.33\textwidth}
    \centering
    \includegraphics[width=\linewidth]{\detokenize{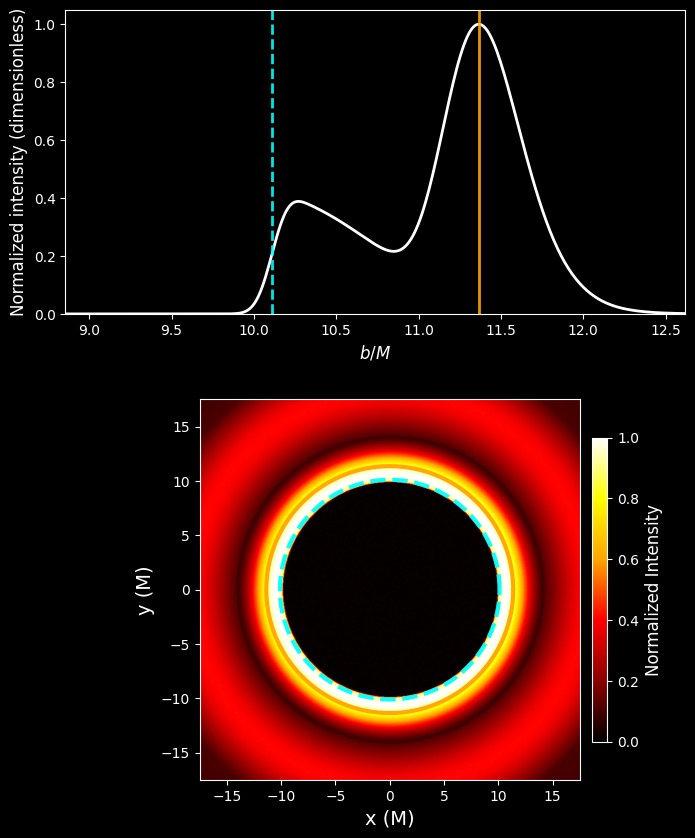}}
    \caption{}
  \end{subfigure}\hfill
  \begin{subfigure}{0.33\textwidth}
    \centering
    \includegraphics[width=\linewidth]{\detokenize{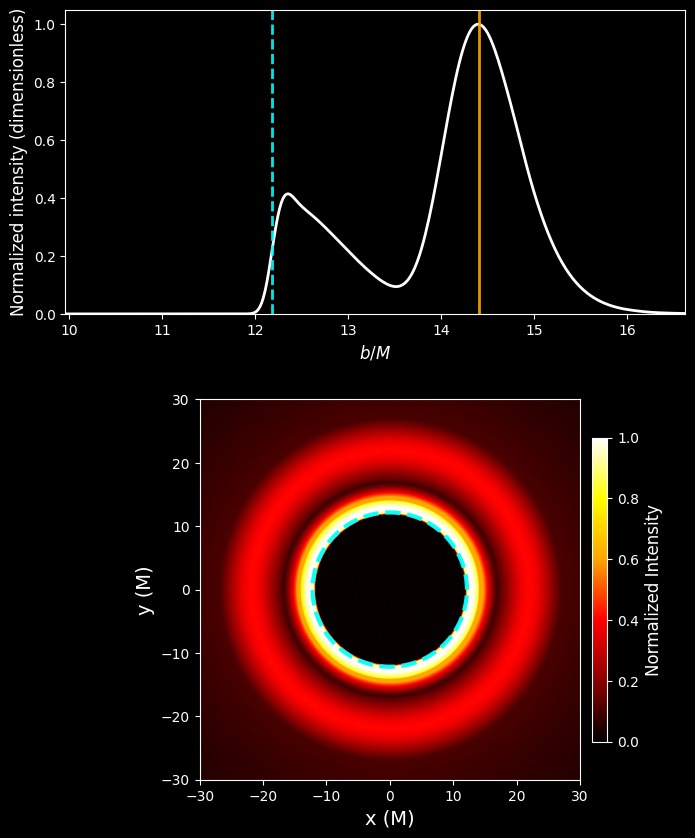}}
    \caption{}
  \end{subfigure}
\caption{Representation of the LV black hole shadows with a global monopole, showing ring radii for three different sets of the parameter values. Top panels display the azimuthally averaged, normalized intensity as a function of $b/M$. The solid (orange) and dashed (cyan) lines indicate the best-fit of the localization of 
the lensing (bright) ring and the photon ring, the critical curve given by $b_c$, respectively, where their maximum brightnesses occur at (a) $b_{\text{peak}}=9.461$ (b) $b_{\text{peak}}=11.364$ and (c) $b_{\text{peak}}=14.000$, while the dashed (critical) lines occur at (a) $b_c =8.665$, (b) $b_c =10.109
$ and (c)  $b_c =12.181
$. Bottom panels display the shadow images themselves for each choice of parameter values:  (a) $\gamma=0.1$, $\kappa\eta^2=0.36$ and $M=1$, (b) $\gamma=0.1$, $\kappa\eta^2=0.423$ and $M=1$ and (c) $\gamma=0.1$, $\kappa\eta^2=0.49$ and $M=1$.  }
  \label{fig:three-panels}
\end{figure}

\begin{figure}[ht!]
  \centering
\begin{subfigure}{0.33\textwidth}
    \centering
    \includegraphics[width=\linewidth]{\detokenize{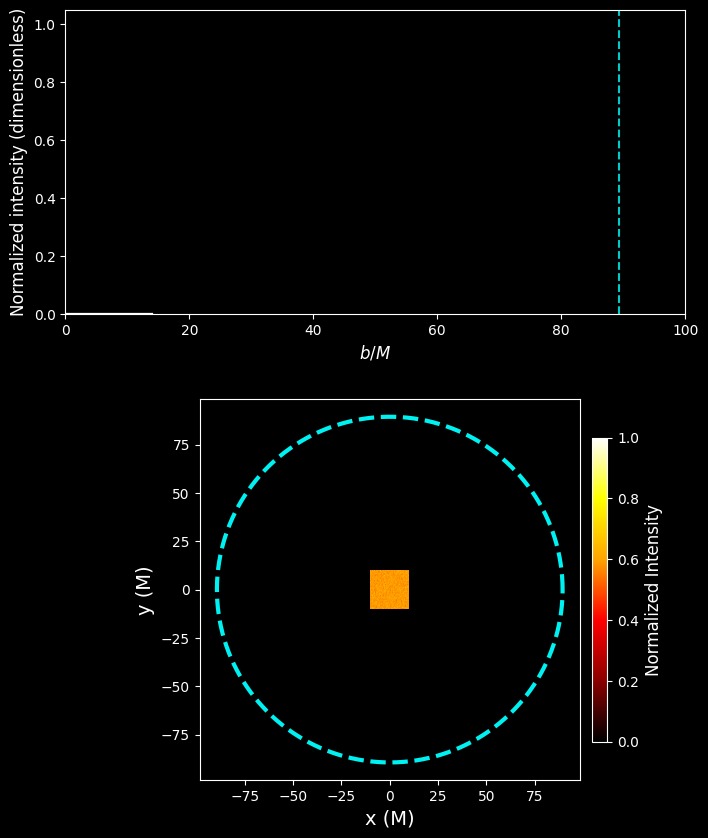}}
    \caption{}
  \end{subfigure}\hfill
  \begin{subfigure}{0.33\textwidth}
    \centering
    \includegraphics[width=\linewidth]{\detokenize{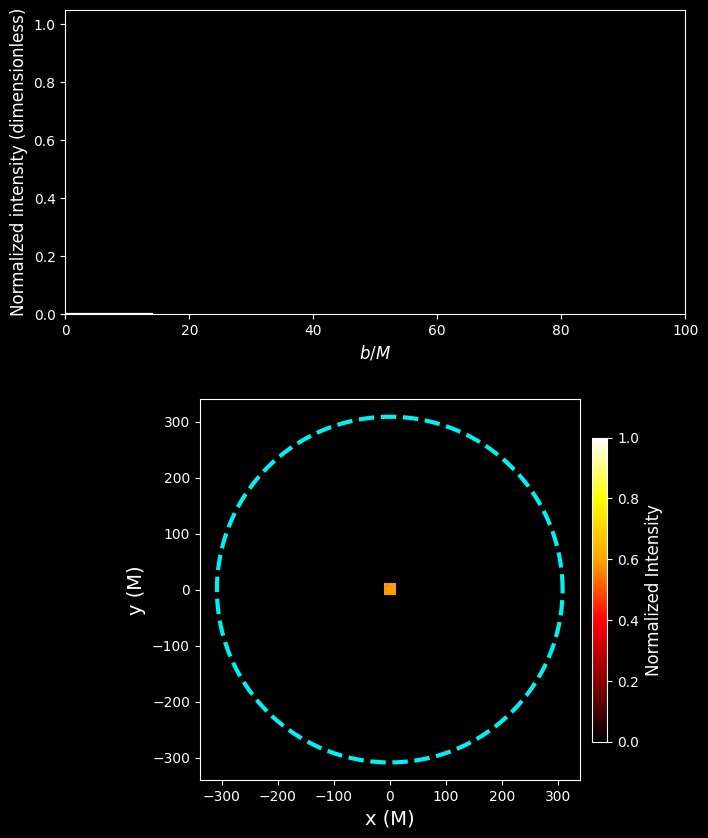}}
    \caption{}
  \end{subfigure}\hfill
  \begin{subfigure}{0.33\textwidth}
    \centering
    \includegraphics[width=\linewidth]{\detokenize{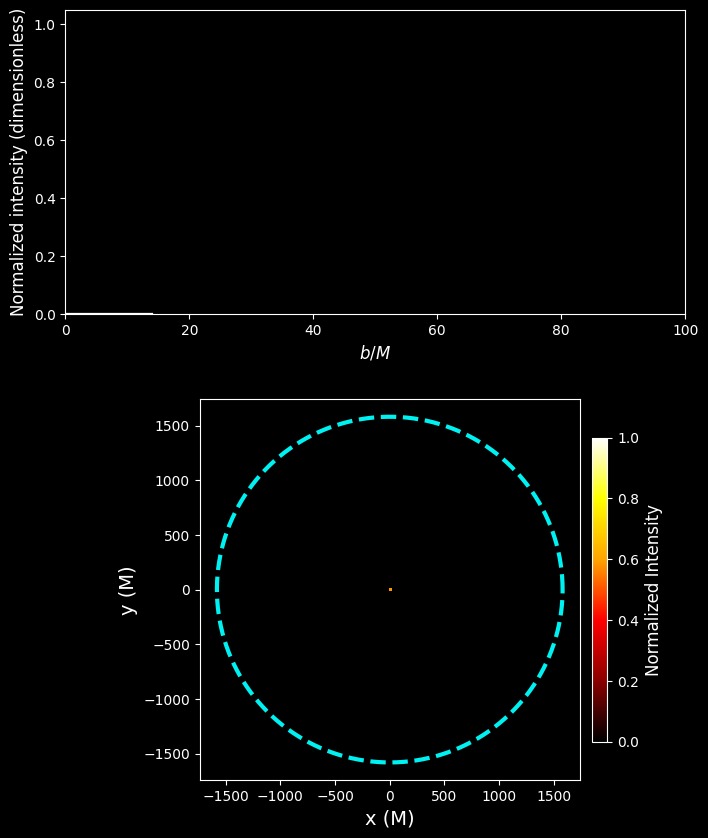}}
    \caption{}
  \end{subfigure}
\caption{Representation of the LV black hole shadows with a global monopole, showing ring radii for three different sets of the parameter values. Top panels display the azimuthally averaged, normalized intensity as a function of $b/M$. Bottom panels (a) $\gamma=0.1$, $\kappa\eta^2=0.94$ and $M=1$, (b) $\gamma=0.1$, $\kappa\eta^2=0.97$ and $M=1$, and (c) $\gamma=0.1$, $\kappa\eta^2=0.99$ and $M=1$.}
  \label{fig:three-panelsB}
\end{figure}

\begin{figure}[ht!]
  \centering
\begin{subfigure}{0.33\textwidth}
    \centering
    \includegraphics[width=\linewidth]{\detokenize{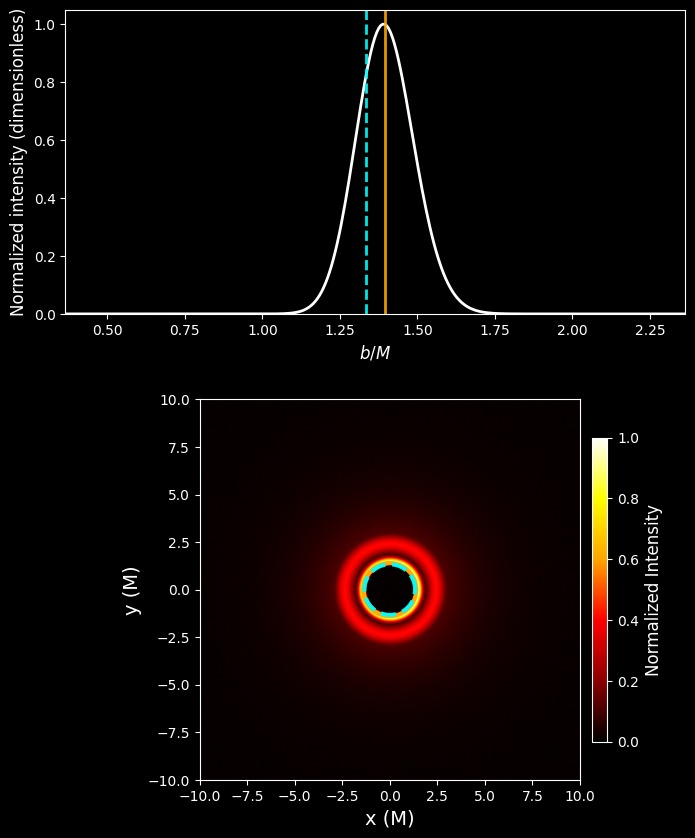}}
    \caption{}
  \end{subfigure}\hfill
  \begin{subfigure}{0.33\textwidth}
    \centering
    \includegraphics[width=\linewidth]{\detokenize{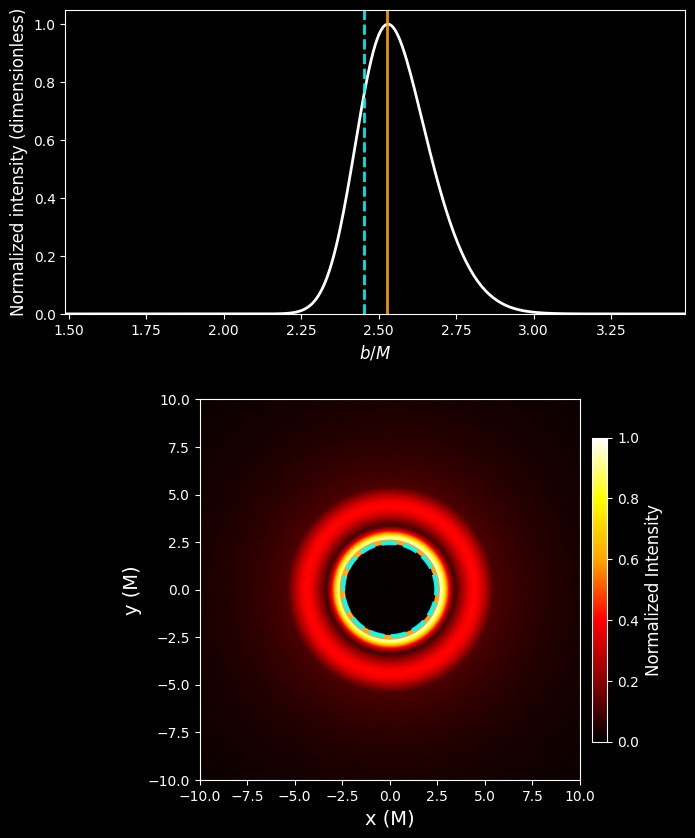}}
    \caption{}
  \end{subfigure}\hfill
  \begin{subfigure}{0.33\textwidth}
    \centering
    \includegraphics[width=\linewidth]{\detokenize{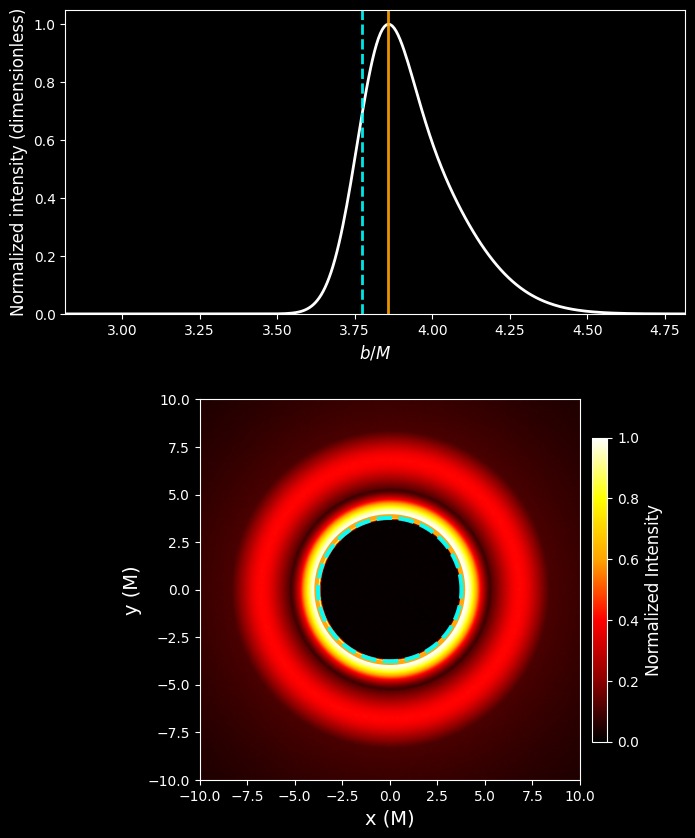}}
    \caption{}
  \end{subfigure}
\caption{Representation of the LV black hole shadows with a global monopole, showing ring radii for three different sets of the parameter values. In the top panels, we have the lensing ring at (a) $b_{\text{peak}}=1.337$, (b) $b_{\text{peak}}=2.464$, and (c) $b_{\text{peak}}=3.833$, and the photon ring (a) $b_{c}=1.335$, (b) $b_{c}=2.452
$, and (c) $b_{\text{peak}}=3.775$. In the bottom panels, we display the shadow images for the following choice of parameters: (a) $\gamma=0.6$, $\kappa\eta^2=0.01$ and $M=1$, (b) $\gamma=0.4$, $\kappa\eta^2=0.01$ and $M=1$ and (c) $\gamma=0.2$, $\kappa\eta^2=0.01$ and $M=1$.  }
  \label{fig:three-panelsC}
\end{figure}

To start, let us define the angular size $\delta$ of the shadow seen by a static observer at area radius $r$,
\begin{equation}
    \delta(r)= \sin^{-1}\left(\frac{R_{\text{opt}}(r_c)}{R_{\text{opt}}(r)}\right)=\sin^{-1}\left(\frac{3M(1-\gamma)}{(1-\kappa\eta^{2})\,r}\,
\sqrt{\,3-\frac{6M(1-\gamma)}{(1-\kappa\eta^{2})\,r}\,}\right),
\end{equation}
where $R_{\text{opt}}(r)=\dfrac{r}{\sqrt{F(r)}}$ is the optical radius -- the area radius that appears in the optical metric related with the LV black hole metric \eqref{metric} and \eqref{sol1} (see f.e. \cite{Cvetic:2016bxi} for more details). Putting all the information together and following the methodology employed in \cite{Gralla, Olmo1, daSilva, Olmo2}, we can construct the shadow images for the LV black hole sourced by a global monopole charge, presented in Fig. \ref{fig:three-panels}, \ref{fig:three-panelsB} and \ref{fig:three-panelsC}. It represents a two-panel figure that jointly describes (i) the top panels describe the brightness distribution around the shadow boundaries and (ii) the bottom panels, the shadow’s size/edge. Let us analyze them individually: as we can be seen from Fig.\ref{fig:three-panels}, the presence of the global charge gives rise to a deformation in the region between the photon ring (dashed line) and the lensing or bright ring (solid line); as $\kappa\eta^2$ grows, the deformation becomes more significant. This is more clear by using the gravitational lensing equations discussed in the previous sections. To the best of our knowledge, let us define the parameter 
\begin{equation}
\chi\equiv \dfrac{1-\kappa\eta^2}{1-\gamma}.
\end{equation}
Notice that the quantities \eqref{bara}, \eqref{barb} and the deflection angle of light \eqref{angle} are modulated by the parameter $\chi$. As a result, the observable parameter which relates the flux of the first image and the total flux of all higher-order images also depends on $\chi$, i.e,
\begin{equation}
    \tilde{r}=e^{2\pi\sqrt{\chi}}.
\end{equation}
Therefore, variations in $\chi$ shift the photon ring by changing the critical curve $b_c$ and reshape the profile between the photon and lensing rings. In particular, the azimuthally averaged intensity (top panels of Fig.~\ref{fig:three-panels}) becomes non-monotonic in the region between $b_c$ and $b_{\text{peak}}$ -- developing a valley -- as $\chi$ decreases (top panels (b) and (c) of Fig.~\ref{fig:three-panels}).

In the limit $\chi\!\to\!\infty$, $r_h$, $b_c$, and $r_s$ collapse to zero; consequently, the gravitational lensing effects disappear (Fig.~\ref{fig:three-panelsB}). This is illustrated for fixed $\gamma=0.1$ while varying $\kappa\eta^2$: bottom panels (a)~$\kappa\eta^2=0.94$, (b)~$0.97$, and (c)~$0.99$.

Figure~\ref{fig:three-panelsC} shows shadow images for fixed global monopole charge $\eta=0.1$ with the Lorentz-violating (LV) parameter varied as follows: bottom panels (a)~$\gamma=0.6$, (b)~$0.4$, and (c)~$0.2$. As $\gamma$ increases, the shadow radius $r_s$ decreases. The top panels exhibit a profile similar to the Schwarzschild case, except that $b_c$ and $b_{\text{peak}}$ are shifted relative to their Schwarzschild values.

\section{Final remarks}\label{s6}

In this paper, one has examined the impact of an antisymmetric self-interacting Kalb–Ramond tensor field with spontaneous Lorentz symmetry breaking on the properties of a black hole with global monopole. For our analysis, we have focused on two important characteristics, fermionic greybody bounds and gravitational lensing in the strong field regime. This research continues the study of implications of Lorentz symmetry breaking in the context of black holes by seeking to fill a gap in the current literature and expand even more our knowledge of LV effects in this context.

Initially, we have investigated the greybody factor of massive and massless spin 1/2 fermions and spin 3/2 fermions (gravitino) by showing the influence of the LV parameter and the global monopole charge. As $\gamma$ increases, the GF decreases and the opposite happens when $\eta$ increases.

The other important topic investigated in this work is gravitational lensing in the strong-field limit. In order to achieve our goal, we have employed the approach developed by Bozza and Tsukamoto \cite{Bozza:2002zj, Tsukamoto:2016jzh}. We applied the strong-field approach to study the shadows cast by the LV black hole sourced by a global monopole produced by optically thin ring/shell emission. In this regard, we found that the shadow properties depend on $\gamma$ and $\eta$. In particular, only $\eta$ modulates the profile of the azimuthally averaged, normalized intensity around the shadows, while both $\gamma$ and $\eta$ control the shadow size. In the limiting case corresponding to $\kappa\eta^2 \to 1$,  gravitational lensing effects vanish.

As a further perspective, we could apply a suitable approach to investigate the greybody factor of vector and tensor perturbations concerning the black hole solution assumed in this work. Another relevant topic of investigation is obtaining the quasinormal modes for our metric.

\section*{Acknowledgments}
\hspace{0.5cm} The authors thank the Funda\c{c}\~{a}o Cearense de Apoio ao Desenvolvimento Cient\'{i}fico e Tecnol\'{o}gico (FUNCAP), the Coordena\c{c}\~{a}o de Aperfei\c{c}oamento de Pessoal de N\'{i}vel Superior (CAPES), and the Conselho Nacional de Desenvolvimento Cient\'{i}fico e Tecnol\'{o}gico (CNPq).  Fernando M. Belchior thanks the Departmento de F\'isica da Universidade Federal da Para\'iba - UFPB for the kind hospitality and has been partially supported by CNPq grant No. 161092/2021-7. Roberto V. Maluf thanks the CNPq for grant no. 200879/2022-7. Ana R. M. Oliveira has been partially supported by CAPES. Albert Yu. Petrov and  Paulo J. Porf\'irio would like to acknowledge the CNPq, respectively for grant No. 303777/2023-0 and grant No. 307628/2022-1.

\end{document}